\documentclass[10pt,journal]{IEEEtran}
\IEEEoverridecommandlockouts 

\usepackage{epsfig}
\usepackage{colordvi}
\usepackage{amssymb,amsfonts,amstext}
\usepackage{amsmath}
\allowdisplaybreaks[4]

\newtheorem{theorem}{Theorem}

\newtheorem{remark}{Remark}

\newcommand {\mc}{\mathcal}

\newcommand{\eqd}{\stackrel{\triangle}{=}}
\newcommand{\dfn}{\stackrel{\triangle}{=}}
\newcommand {\exe} {\stackrel{\cdot} {=}}

\newcommand {\pr} {\mbox{Pr}}

\newcommand {\bv} {\mbox{\boldmath $v$}}

\newcommand {\bx} {\mbox{\boldmath $x$}}
\newcommand {\by} {\mbox{\boldmath $y$}}
\newcommand {\bz} {\mbox{\boldmath $z$}}

\newcommand {\bE} {\mbox{\boldmath $E$}}

\newcommand {\bX} {\mbox{\boldmath $X$}}
\newcommand {\bY} {\mbox{\boldmath $Y$}}

\newcommand{\calC}{{\cal C}}

\newcommand{\calS}{{\cal S}}
\newcommand{\calT}{{\cal T}}

\newcommand{\calV}{{\cal V}}

\newcommand{\calX}{{\cal X}}
\newcommand{\calY}{{\cal Y}}

\newcommand{\beq}{\begin{equation}}
\newcommand{\eeq}{\end{equation}}

\newcommand{\argmax}{\operatornamewithlimits{arg\,max}}
\newcommand{\argmin}{\operatornamewithlimits{arg\,min}}

\begin{document}

\newcommand{\tp}{\ensuremath{P_{\hat{X}_1\hat{X}_2\hat{Y}_1}}}
\newcommand{\tpp}{\ensuremath{P_{\hat{X}_1'\hat{X}_2'\hat{Y}_1'}}}
\newcommand{\ph}{\ensuremath{\hat{P}}}
\newcommand{\php}{\ensuremath{\hat{P}'}}
\newcommand{\Iyxz}{\ensuremath{I(\hat{X}_2;\hat{X}_1,\hat{Y}_1)}}
\newcommand{\Iyxzp}{\ensuremath{I(\hat{X}_2';\hat{X}_1',\hat{Y}_1')}}
\newcommand{\Iyz}{\ensuremath{I(\hat{X}_2;\hat{Y}_1)}}
\newcommand{\Iyzp}{\ensuremath{I(\hat{X}_2';\hat{Y}_1')}}
\newcommand{\Ixz}{\ensuremath{I(\hat{X}_1;\hat{Y}_1)}}
\newcommand{\Ixzp}{\ensuremath{I(\hat{X}_1';\hat{Y}_1')}}
\newcommand{\elq}{\ensuremath{\bE_{\hat{X}_1\hat{X}_2\hat{Y}_1}\left[\log q_1\left(\hat{Y}_1|\hat{X}_1,\hat{X}_2\right)\right]}}
\newcommand{\elqp}{\ensuremath{\bE_{\hat{X}_1'\hat{X}_2',\hat{Y}_1'}\left[\log q_1\left(\hat{Y}_1'|\hat{X}_1',\hat{X}_2'\right)\right]}}
\newcommand{\qxa}{\ensuremath{Q_1}}
\newcommand{\qxb}{\ensuremath{Q_2}}

% used by alternative expression 
\newcommand{\Ixyz}{\ensuremath{I(\hat{X}_1;\hat{X}_2,\hat{Y}_1)}}
\newcommand{\Ixyzp}{\ensuremath{I(\hat{X}_1';\hat{X}_2',\hat{Y}_1')}}
\newcommand{\Hzx}{\ensuremath{H(\hat{Y}_1|\hat{X}_1)}}
\newcommand{\Hzy}{\ensuremath{H(\hat{Y}_1|\hat{X}_2)}}
\newcommand{\Hzxy}{\ensuremath{H(\hat{Y}_1|\hat{X}_1,\hat{X}_2)}}
\newcommand{\Dpzxyq}{\ensuremath{D(P_{\hat{Y}_1|\hat{X}_1\hat{X}_2}||q_1|
P_{\hat{X}_1\hat{X}_2})}}
\newcommand{\Da}[1]{\ensuremath{D(P_{\hat{Y}^{(#1)}_1|\hat{X}^{(#1)}_1\hat{X}^{(#1)}_2}||q_1|
P_{\hat{X}^{(#1)}_1\hat{X}^{(#1)}_2})}}
\newcommand{\Dpa}[1]{\ensuremath{D(P_{\hat{Y}^{'(#1)}_1|\hat{X}^{'(#1)}_1\hat{X}^{'(#1)}_2}||q_1|
P_{\hat{X}^{'(#1)}_1\hat{X}^{'(#1)}_2})}}
\newcommand{\tpa}[1]{\ensuremath{P_{\hat{X}^{(#1)}_1\hat{X}^{(#1)}_2\hat{Y}^{(#1)}_1}}}
\newcommand{\tppa}[1]{\ensuremath{P_{\hat{X}^{'(#1)}_1\hat{X}^{'(#1)}_2\hat{Y}^{'(#1)}_1}}}
\newcommand{\Iyxza}[1]{\ensuremath{I(\hat{X}^{(#1)}_2;\hat{X}^{(#1)}_1,\hat{Y}^{(#1)}_1)}}
\newcommand{\Iyxzpa}[1]{\ensuremath{I(\hat{X}^{'(#1)}_2;\hat{X}^{'(#1)}_1,\hat{Y}^{'(#1)}_1)}}
\newcommand{\Ixyzpa}[1]{\ensuremath{I(\hat{X}^{'(#1)}_1;\hat{X}^{'(#1)}_2,\hat{Y}^{'(#1)}_1)}}
\newcommand{\Ixyza}[1]{\ensuremath{I(\hat{X}^{(#1)}_1;\hat{X}^{(#1)}_2,\hat{Y}^{(#1)}_1)}}
\newcommand{\Iyza}[1]{\ensuremath{I(\hat{X}^{(#1)}_2;\hat{Y}^{(#1)}_1)}}
\newcommand{\Iyzpa}[1]{\ensuremath{I(\hat{X}^{'(#1)}_2;\hat{Y}^{'(#1)}_1)}}
\newcommand{\Ixza}[1]{\ensuremath{I(\hat{X}^{(#1)}_1;\hat{Y}^{(#1)}_1)}}
\newcommand{\Ixzpa}[1]{\ensuremath{I(\hat{X}^{'(#1)}_1;\hat{Y}^{'(#1)}_1)}}
\newcommand{\Hzxa}[1]{\ensuremath{H(\hat{Y}^{(#1)}_1|\hat{X}^{(#1)}_1)}}
\newcommand{\elqa}[1]{\ensuremath{E\left[\log q_1(\hat{Y}^{(#1)}_1|\hat{X}^{(#1)}_1,\hat{X}^{(#1)}_2)\right]}}
\newcommand{\elqpa}[1]{\ensuremath{E\left[\log
       q_1(\hat{Y}^{'(#1)}_1|\hat{X}^{'(#1)}_1,\hat{X}^{'(#1)}_2)\right]}}
\newcommand{\Ixya}[1]{\ensuremath{I(\hat{X}^{(#1)}_1;\hat{X}^{(#1)}_2)}}
\newcommand{\Ixy}{\ensuremath{I(\hat{X}_1;\hat{X}_2)}}
\newcommand{\Ixypa}[1]{\ensuremath{I(\hat{X}^{'(#1)}_1;\hat{X}^{'(#1)}_2)}}

\newcommand{\Ixzgya}[1]{\ensuremath{I(\hat{X}^{(#1)}_1;\hat{Y}^{(#1)}_1|\hat{X}^{(#1)}_2)}}

\title{Error Exponents of Optimum Decoding for the Interference Channel}
  
%%%%%%%%%%%%%%%%%%%%
\author{Raul H. Etkin$^\dag$,~\IEEEmembership{Member,~IEEE,}
Neri Merhav$^\ddag$,~\IEEEmembership{Fellow,~IEEE,}
and Erik Ordentlich$^\dag$,~\IEEEmembership{Senior Member,~IEEE}%
\\
raul.etkin@hp.com, merhav@ee.technion.ac.il, erik.ordentlich@hp.com

\thanks{$^\dag$R. Etkin and E. Ordentlich are with Hewlett-Packard Laboratories, Palo Alto, CA 94304, USA.}
\thanks{$^\ddag$N. Merhav is with Technion - Israel Institute of Technology, Haifa 32000, Israel. Part of this work was done while N.~Merhav was visiting Hewlett--Packard Laboratories
in the Summers of 2007 and 2008.}
}
%%%%%%%%%%%%%%%%%%%% 
\maketitle
\begin{abstract}
Exponential error bounds for the finite--alphabet interference channel (IFC) with two transmitter--receiver pairs, are investigated under the random coding regime. Our focus is on optimum decoding, as opposed to
heuristic decoding rules that have been used in previous works, like joint typicality
decoding, decoding based on interference cancellation, and decoding that considers the
interference as additional noise. Indeed, the fact that the actual
interfering signal is a codeword and not an i.i.d. noise process complicates the application of conventional techniques to the performance analysis of the optimum decoder.
Using analytical tools rooted in
statistical physics, we derive a single letter expression 
for error exponents achievable under optimum decoding 
and demonstrate strict improvement over error exponents
obtainable using suboptimal decoding rules, but which are amenable
to more conventional analysis.
\end{abstract}


\section{Introduction}
\label{sec:intro}
The $ M $-user interference channel (IFC) models the communication
between $ M $ transmitter-receiver pairs, wherein each receiver 
must 
decode its corresponding transmitter's message from a signal that is corrupted by interference from the
other transmitters, in addition to channel noise.  The information
theoretic analysis of the IFC was initiated over 30 year ago
and has recently witnessed a resurgence of interest, motivated by new
potential applications, such as wireless communication over
unregulated spectrum.

Previous work on the IFC has focused on obtaining inner and outer
bounds to the capacity region for memoryless interference and noise,
with a precise characterization of the capacity region remaining
elusive for most channels, even for $ M = 2 $ users. The best known
inner bound for the IFC is the Han-Kobayashi (HK) region, established
in~\cite{HK81}.  It has been found to be tight in certain special cases
(\cite{HK81,Car75}), and recently was found to be tight to within 1 bit for the
two user Gaussian IFC~\cite{ETW06}.  No achievable rates that lie outside the HK
 region are known for any IFC with $M=2$ users.

Our aim in this paper is to extend the study of achievable schemes to
the analysis of error exponents, or exponential rates of decay of
error probabilities, that are attainable as a function of user rates.
To our knowledge, there has been no prior treatment of error exponents
for the IFC.  In particular, the error bounds underlying the achievability results
in~\cite{HK81} yield vanishing error exponents (though still decaying
error probability) at all rates.

The notion of an error exponent region, or a set of achievable
exponential rates of decay in the error probabilities for different
users at a given operating rate-tuple in a multi-user communication
network, was formalized recently in~\cite{Weng+08}, and studied therein
for Gaussian multiple access and broadcast channels.    
Our main result, presented in Section~\ref{sec:main}, is a single letter
characterization of an achievable error exponent region, as a function
of user rates, for the $ M = 2 $ user finite alphabet, memoryless
interference channel.  
The region is derived by bounding the average
error probability of random codebooks comprised of i.i.d.\ codewords
uniformly distributed over a type class, under maximum likelihood (ML)
decoding at each user. Unlike the single user setting, in this case,
the effective channel determining each receiver's ML
decoding rule is induced both by the noise {\em and} the interfering
user's codebook.  Our focus on optimal decoding is a departure from
the conventional achievability arguments in~\cite{HK81} and elsewhere,
which are based on joint-typicality decoding, with restrictions on the
decoder to ``treat interference as noise'' or to ``decode the
interference'' in part or in whole.  However, in this work, we confine our analysis to  
codebook ensembles that are simpler than the superposition codebooks of~\cite{HK81}. 

The analysis of the probability of decoding error under optimal decoding 
is complicated due to correlations induced by the interfering signal. 
Usual methods for bounding the probability of error based on 
Jensen's inequality and
other related inequalities (see, e.g., (\ref{distexp}) below) 
fail to give good results. 
Our bounding approach combines some of the
classical information theoretic approaches of \cite{Gal68} 
and \cite{CK81} with an analytical technique from statistical
physics that was applied recently to the study of single user channels
in~\cite{Mer07,Mer08}.  More specifically,
as in \cite{Gal68}, we use auxiliary parameters $\rho$ and
$\lambda$ to get an upper bound on the average probability of decoding
error under ML decoding, which we then bound using the method of types
\cite{CK81}. Key in our derivation is the use of distance enumerators
in the spirit of \cite{Mer07} and \cite{Mer08}, 
which allows us to avoid using Jensen's 
inequality in some steps, and allows us to maintain exponential
tightness in other inequalities by applying them to only 
polynomially few terms (as opposed to exponentially many)
in certain sums that bound the probability of decoding error.
It should be emphasized, in this context, that the use of this technique was pivotal to
our results. Our earlier attempts, that were based on more `traditional'
tools, failed to provide meaningful results. In fact, they all turned
out to be inferior to some trivial bounds.

The paper is organized as follows. The notation, various definitions, and the
channel model assumed throughout the paper are detailed in
Section~\ref{sec:notation}.  In Section~\ref{sec:background},
we derive an ``easy''
set of attainable error exponents which
we shall treat as a benchmark for the exponents of the main section,
Section~\ref{sec:main}.  The ``easy'' exponents are obtained by simple
extensions  
to the interference channel of existing error exponent 
results for single user and multiple access channels, based on random constant 
composition codebooks and suboptimal decoders.
Then, in Section~\ref{sec:main}, we derive another set of attainable
exponents by analyzing ML decoding for the
channel induced by the interfering codebook. In Section \ref{sec:convopt},
we show that the minimizations required to evaluate the new error exponents can be
written as convex optimization problems, and, as a result, can be solved efficiently.
We follow this up in
Section~\ref{sec:results} with a numerical comparison
of the new exponents with the baseline exponents of Section \ref{sec:background} for a simple
IFC.  These numerical results demonstrate that the new exponents are never worse (at least for the chosen channel and parameters) and, for most rates, strictly improve over the baseline exponents.

An earlier version of this work was presented in \cite{EMO08}.

\section{Notation, Definitions, and Channel Model}
\label{sec:notation}
Unless otherwise stated, we use lowercase and uppercase letters for scalars, boldface 
lowercase letters for vectors, uppercase (boldface) letters for random variables (vectors), 
and calligraphic letters for sets. For example, $a$ is a scalar, $\bv$ is a vector,
$X$ is a random variable, $\bX$ is a random vector, and $\calS$ is a set.  For a real number $ a $ we shall, on occasion, let $ \overline{a} $ denote $ 1 - a $.
Also, we use $\log(\cdot)$ to denote natural logarithm, $\bE$ to denote 
expectation, and $\pr$ to denote probability. 
For independent random variables $X$ and $Y$ 
distributed according to $P_{X,Y}(x,y)=P_X(x) P_Y(y)$, 
$(x,y)\in \calX \times \calY$, we denote the conditional expectation operator 
$\bE_X(\cdot)$ as $\bE_X(f(X,Y))\dfn \sum_{x\in\calX} f(x,Y) P_X(x)$ for any function $f(\cdot,\cdot)$. All information quantities (entropy, mutual information, etc.) and rates are in nats. 
Finally, we use $\doteq$, $\stackrel{.}{\le}$, etc., to denote equality or inequality to the first order in the exponent, i.e. $a_n \doteq b_n \Leftrightarrow \lim_{n\to \infty} \frac{1}{n} \log \frac{a_n}{b_n} = 0$; $a_n \stackrel{.}{\le} b_n \Leftrightarrow \lim \sup_{n\to \infty} \frac{1}{n} \log \frac{a_n}{b_n} \le 0$.

The empirical probability mass function of the finite alphabet sequence 
$\bv=(v(1),\ldots,v(n))$ with alphabet $\calV$
is denoted as the vector $\{P_{\bv}(v),~v\in\calV\}$, where each $P_{\bv}(v)$
is the relative frequency of $v(i)=v$ along $\bv$. The type class associated
with an empirical probability mass function $P$, which will be denoted
by $\calT_P$, is the set of all $n$--vectors $\{\bv\}$ whose empirical
probability mass function is equal to $P$.
Similar conventions will apply to pairs and triples of vectors
of length $n$, which are defined over the corresponding
product alphabets. Information measures pertaining to empirical distributions will be denoted using the standard notational conventions, except that we use ``$\hat\quad$'' as well as subscripts that indicate the sequences from which these empirical distributions were extracted. For example, we write
 $\hat{H}_{\bx \by \bz}(X,Y|Z)$ and $\hat{I}_{\bx \by \bz}(X,Y;Z)$ to denote the conditional entropy of $(X,Y)$ given $Z$ and the mutual information between $(X,Y)$ and $Z$, respectively, computed with respect to the empirical distribution $P_{\bx \by \bz}(x,y,z)$. We denote the relative entropy or Kullback-Leibler divergence between distributions $P_X$ and $P_Y$ as $D(P_X||P_Y)\dfn\sum_{x} P_X(x) \log (P_X(x)/P_Y(x))$, and we write $D(P_{X|Z}||P_{Y|Z}|P_Z)$ for the conditional relative entropy between conditional distributions $P_{X|Z}$ and $P_{Y|Z}$ conditioned on $P_Z$, which is defined as $D(P_{X|Z}||P_{Y|Z}|P_Z)\dfn \sum_{x,z} P_Z(z) P_{X|Z}(x|z) \log(P_{X|Z}(x|z)/P_{Y|Z}(x|z))$ .

We continue with a formal description of the two--user IFC setting.
Let $\bx_i = (x_{i}(1),\ldots,x_{i}(n)) \in\calX^n_i$, $ i = 1, 2 $,
denote the channel input signals of the two transmitters, and let
$\by_i = (y_{i}(1),\ldots,y_{i}(n)) \in\calY^n_i$ be the corresponding
channel outputs received by decoders 1 and 2, where $ \calX_i $ and $
\calY_i $ denote the input and output alphabets, and which we assume
to be finite.  Each (random) output symbol pair $ (Y_{1}(j),Y_{2}(j)) $
is assumed to be conditionally independent of all other outputs, and
all input symbols, given the two corresponding (random) input symbols
$ (X_{1}(j),X_{2}(j)) $, and the corresponding conditional probability
is assumed to be constant from symbol to symbol.  An $ (n,R_1,R_2) $
code for the IFC consists of pairs of encoding and
decoding functions, $ (f_1, f_2) $ and $ (g_1, g_2) $, respectively,
where $ f_i:\{1,\ldots,M_i\}
\rightarrow \calX^n_i $,
$M_i=\lceil e^{nR_i}\rceil$, and
$ g_i:\calY^n_i \rightarrow \{1,\ldots,M_i\} $, $i=1,2$.  The
performance of the code is characterized by a pair of error
probabilities $ P_{e,i} = \pr(\hat{W}_i \neq W_i) $, $ i = 1, 2$, where
$ \hat{W}_i = g_i(\bY_i) $ and $ \bY_i $ is the random output when
user $ i $ transmits $ \bX_i = f_i(W_i) $, 
assuming the messages $ W_i $ are uniformly distributed on the
sets of indices $ \{1,2,\ldots,M_i\}$, $ i = 1, 2
$.  The per 
user error probabilities depend on the channel only through the
marginal conditional distributions of the channel outputs given the
corresponding channel input pairs.  We shall denote these conditional
distributions as $ q_i(y|x_1,x_2)
\stackrel{\triangle}{=} \pr(Y_{i}(j) = y | (X_{1}(j),X_{2}(j)) = (x_1,x_2))$.

A pair of error exponents $ (E_1,E_2) $ is attainable at a rate pair $
(R_1,R_2) $ if there is a sequence of $ (n, R_1, R_2) $ codes
satisfying $ E_i \leq \liminf_{n\to\infty} -(1/n)\log P_{e,i} $ for $ i = 1, 2 $.
The set of all attainable error exponents at $ (R_1,R_2) $ comprises
the error exponent region at $ (R_1,R_2) $ and we shall denote it as $
{\cal E}(R_1,R_2) $.  The main result of this paper is a single letter
characterization of a non--trivial subset of $ {\cal E}(R_1,R_2) $ for
each $ R_1, R_2 $.

\section{Background}
\label{sec:background}
In this section, we present achievable error exponents for the interference channel which are based on known results of error exponents for single user and multiple access channels (MAC) for fixed composition codebooks \cite{PW85, LH96, Cha08}. These exponents will be used as a baseline for comparing the performance of the error exponents that we derive in Section \ref{sec:main}. 

In the following, we will focus on the error performance of user 1, and as a result, all explanations and expressions will be specialized to receiver 1. Similar expressions also hold for user 2 with the exchange of indices $1 \leftrightarrow 2$.

A possibly suboptimal decoder for the interference channel can be obtained from a given multiple access channel decoder by simply ignoring the decoded message of the interfering transmitter. For example, following \cite{LH96}, we can use a minimum entropy decoder that for a given received vector $\by_1$ at receiver $1$ computes $(\hat{\bx}_1,  \hat{\bx}_2)$
\[
(\hat{\bx}_1,  \hat{\bx}_2) = \argmin_{(\tilde{\bx}_1,\tilde{\bx}_2) \in \calC_1\times \calC_2}  \hat{H}_{\tilde\bx_1 \tilde\bx_2 \by_1}(X_1,X_2|Y_1),
\]
and throws away $\hat{\bx}_2$.

It follows from \cite{LH96} that for random codebooks of fixed composition $Q_1, Q_2$, the average probability of decoding both messages in error, where the averaging is done over the random choice of codebooks, satisfies: 
\[
\Pr(\hat{\bx}_1\ne \bx_1, \hat{\bx}_2\ne \bx_2) \stackrel{.}{\le}e^{-n E_{1,2}}
\]
where
\begin{align}
E_{1,2}=&\min_{\tp: P_{\hat{X}_1}=Q_1, P_{\hat{X}_2}=Q_2} D(P_{\hat{Y}_1|\hat{X}_1 \hat{X}_2}|| q_1| P_{\hat{X}_1,\hat{X}_2})\nonumber\\
&+ \Ixy\nonumber\\
&+|I(\hat{X}_1;\hat{Y}_1)+I(\hat{X}_2;\hat{X}_1,\hat{Y}_1)-R_1-R_2|^+\nonumber
\end{align}
with $|\cdot|^+ = \max\{\cdot, 0\}$.

In addition, the average probability of decoding the message of the interfering transmitter correctly but the message of the desired transmitter incorrectly satisfies:
\[
\Pr(\hat{\bx}_1\ne \bx_1, \hat{\bx}_2= \bx_2) \stackrel{.}{\le}e^{-n E_{1|2}}
\]
where
\begin{align}
E_{1|2}=&\min_{\tp: P_{\hat{X}_1}=Q_1, P_{\hat{X}_2}=Q_2} D(P_{\hat{Y}_1|\hat{X}_1 \hat{X}_2}|| q_1| P_{\hat{X}_1,\hat{X}_2})\nonumber\\
&+ \Ixy+|I(\hat{X}_1;\hat{X}_2,\hat{Y}_1)-R_1|^+.\nonumber
\end{align}
Therefore, the overall average error performance of this MAC decoder in the IFC satisfies:
\[
\Pr(\hat{\bx}_1\ne \bx_1) \stackrel{.}{\le}e^{-n \min\{E_{1,2},E_{1|2}\}}.
\]

A second suboptimal decoder that leads to tractable error performance bounds is the single user maximum mutual information decoder (which in this case coincides with the minimum entropy decoder):
\[
\hat{\bx}_1 = \argmax_{\bx_1 \in \calC_1} \hat{I}_{\bx_1 \by_1}(X_1;Y_1).
\]
In this case, standard application of the method of types \cite{Cha08} leads to the following bound on the average error probability under random fixed composition codebooks of types $Q_1, Q_2$:
\[
\Pr(\hat{\bx}_1\ne \bx_1) \stackrel{.}{\le}e^{-n E_{1}},
\]
where
\begin{align}
E_{1}=&\min_{\tp: P_{\hat{X}_1}=Q_1, P_{\hat{X}_2}=Q_2} D(P_{\hat{Y}_1|\hat{X}_1 \hat{X}_2}|| q_1| P_{\hat{X}_1,\hat{X}_2})\nonumber\\
&+ \Ixy+|I(\hat{X}_1;\hat{Y}_1)-R_1|^+.\nonumber
\end{align}

We can choose the better decoder between these two, that leads to the better error performance. Therefore, we obtain that
\beq
E_{B,1}=\max\{E_1;\min\{E_{1,2};E_{1|2}\}\}
\eeq
is an achievable error exponent at receiver 1, with an analogous exponent following for receiver 2.

\section{Main Result}
\label{sec:main}
Our main contribution is stated in the following theorem,
which presents a new error exponent region 
for the discrete memoryless  two--user IFC.
While the full proof appears in Appendix~\ref{sec:proof}, we also provide
a proof outline below, to give an idea of the main steps.
\begin{theorem}
\label{thm:1}
For a discrete memoryless two-user IFC as defined in Section~\ref{sec:intro}, for a family of block codes of rates $R_1$ and $R_2$ a decoding error probability for user 1 satisfying
\beq
\liminf_{n\to\infty} -\frac{1}{n} \log \overline{P}_{e,1}(n) \ge E_{R,1}(R_1,R_2,\qxa, \qxb,\rho,\lambda)
\label{eqn:pe}
\eeq
can be achieved as the block length 
of the codes $n$ goes to infinity, where the 
error exponent $E_{R,1}(R_1,R_2,\qxa, \qxb,\rho,\lambda)$ is given by
\begin{align}
&E_{R,1}(R_1,R_2,\qxa, \qxb,\rho,\lambda)=
\Bigg\{
R_2-\rho R_1 + \min \Bigg\{\nonumber\\
& \min_{\substack{(\tp,\tpp)\\
\in \calS_1(Q_1,Q_2)}}
f_1\left(\rho,\lambda,\tp,\tpp\right);\nonumber\\
&\min_{\substack{(\tp,\tpp)\\
\in \calS_2(Q_1,Q_2,R_2)}} 
 f_2\left(\rho,\lambda,\tp,\tpp\right)\Bigg\}\Bigg\}
\label{eqn:erx}
\end{align}
where
\begin{align}
f_1 \dfn & g(\rho,\lambda,\tp,\tpp) - H(\hat{Y}_1|\hat{X}_1)+\rho I(\hat{X}_1';\hat{Y}_1') \nonumber\\
&+ \max\bigg\{ \Iyxz-R_2; \nonumber\\
&\quad \quad \quad \quad \quad \quad \quad \quad \overline{\rho\lambda}(\Iyxz-R_2)\bigg\}\nonumber\\
&+ \max\bigg\{\overline{\rho}I(\hat{X}_2';\hat{Y}_1') +\rho \Iyxzp-R_2;\nonumber\\
&\rho (\Iyxzp-R_2); \rho \lambda(\Iyxzp-R_2)\bigg\} \label{eqn:f1}\\
f_2 \dfn & g(\rho,\lambda,\tp,\tpp) -H(\hat{Y}_1|\hat{X}_1)\nonumber\\
& +\rho I(\hat{X}_1';\hat{X}_2',\hat{Y}_1') +\Iyxz - R_2 \label{eqn:f2}
\end{align}
with
\begin{align}
g \dfn&  -\overline{\rho\lambda}E_{\hat{X}_1,\hat{X}_2,\hat{Y}_1}\log q_1(\hat{Y}_1|\hat{X}_1,\hat{X}_2)\nonumber\\
& \quad\quad\quad\quad -\rho\lambda E_{\hat{X}_1',\hat{X}_2',\hat{Y}_1'}\log q_1(\hat{Y}_1'|\hat{X}_1',\hat{X}_2')\nonumber
\end{align}
\end{theorem}
and
\begin{align}
\calS_1(Q_1,Q_2)\dfn&\big\{(\tp,\tpp) \in \calS^2: P_{\hat{Y}_1}=P_{\hat{Y}_1'}, \nonumber\\
& P_{\hat{X}_1}=P_{\hat{X}_1'}=\qxa, P_{\hat{X}_2}=P_{\hat{X}_2'}=\qxb  \big\}
\label{eqn:s1}
\end{align}
\begin{align}
\calS_2(Q_1,Q_2,R_2)\dfn&\big\{(\tp,\tpp) \in \calS^2: \nonumber\\
& P_{\hat{X}_1}=P_{\hat{X}_1'}=\qxa, P_{\hat{X}_2}=P_{\hat{X}_2'}=\qxb, \nonumber\\
& R_2 \le I(\hat{X}_2;\hat{Y}_1), P_{\hat{X}_2,\hat{Y}_1}=P_{\hat{X}_2',\hat{Y}_1'} \big\}
\label{eqn:s2}
\end{align}
where $\calS$ is the probability simplex in $\mc{X}_1\times \mc{X}_2\times \mc{Y}_1$. In the bound (\ref{eqn:pe}),  $(\rho,\lambda) \in [0,1]^2$ can be chosen to maximize the error exponent $E_{R,1}$. 

In eqs. (\ref{eqn:pe}), (\ref{eqn:erx}), (\ref{eqn:s1}), and (\ref{eqn:s2}), $\qxa$ and $\qxb$ are probability distributions defined over the alphabets $\mc{X}_1$ and $\mc{X}_2$ respectively. 

Expressions for the error probability $P_{e,2}$ and error exponent $E_{R,2}$ equivalent to (\ref{eqn:pe}) and (\ref{eqn:erx}) can be stated for the receiver of user 2 by replacing $X_1 \leftrightarrow X_2$, $Y_1 \rightarrow Y_2$, and $q_1\rightarrow q_2$ in all the expressions. By varying $\qxa$ and $\qxb$ over all probability distributions in $\mc{X}_1$ and $\mc{X}_2$ respectively, we obtain the error exponent region for fixed rates $R_1$ and $R_2$.

\begin{remark} A lower bound to $ E_{R,1}^* \stackrel{\triangle}{=} \max_{\rho,\lambda}E_{R,1}(R_1,$ $R_2,\qxa, \qxb,\rho,\lambda) $ is
derived in Appendix~\ref{sec:altexpr} (cf. equation (\ref{eq:altexprfinal})) that is closer in form to the
expressions underlying the benchmark exponent $ E_{B,1} $ presented
above.  In particular, this lower bound allows us to establish
analytically (see Appendix~\ref{sec:altexpr}) that $ E_{B,1} \leq E_{R,1}^* $ at $ R_1 = 0 $ (and for
sufficiently small $ R_1 $).  Numerical computations, as presented in Section \ref{sec:results},
indicate that this inequality can be strict. 

A second application of the lower bound (\ref{eq:altexprfinal}) is to determine the set
of rate pairs $R_1, R_2$ for which $E_{R,1}^*>0$. We show in Appendix~\ref{sec:altexpr}
that this region includes
\begin{multline*}
\mc{R}_1=\{ R_1 < \Ixz \} \cup \Big\{\{R_1 + R_2 <
I(\hat{Y}_1;\hat{X}_1,\hat{X}_2) 
\} \\ \cap \{R_1 < I(\hat{X}_1;\hat{Y}_1|\hat{X}_2)\Big\}, 
\end{multline*}
with an analogous region following for the set where $E_{R,2}^*>0$ (see Fig. \ref{fig:rateregion}).

\begin{figure}[htb]
\centerline{ \psfig{figure=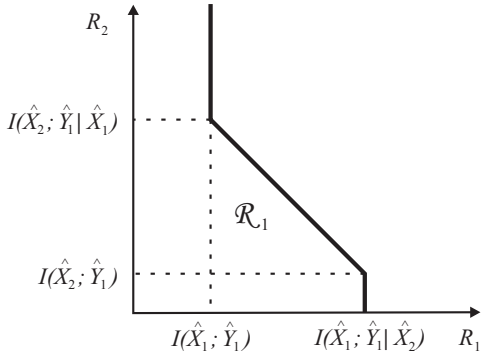,,width=2.5in}}
\vspace{-0.1in}
\caption{\footnotesize Rate region $\mc{R}_1$ where $E_{R,1}^*>0$.}
\label{fig:rateregion} 
\end{figure}
\vspace{0.05in}

Furthermore, it is shown in \cite{Cha08} that the error exponent 
achievable for user no. 1 with optimal decoding and random fixed composition codebooks 
is zero outside the closure of the region $\mc{R}_1$. This result, together with our contribution
characterize the rate region where the attainable exponents with random constant composition codebooks are positive. Finally, it can be shown that this region is contained in the HK region~\cite{Cha08}.
\end{remark}

\begin{remark}
Theorem \ref{thm:1} presents an asymptotic upper bound on the average probability of decoding error for fixed composition codebooks, where the averaging is done over the random choice of codebooks. It is straightforward to show (see, e.g., \cite{Weng+08}) that there exists a specific (i.e. non-random) sequence of fixed composition codebooks of increasing block length $n$ for which the same asymptotic error performance can be achieved.
\end{remark}
\vspace{0.5cm}

\noindent
{\it Proof Outline.}
For $n$ non--negative reals $a_1,\ldots,a_n$ and $b\in[0,1]$,
the following inequality \cite[Problem 4.15(f)]{Gal68} will be frequently used: 
\begin{equation}
\label{distexp}
\left(\sum_{i=1}^na_i\right)^b\le \sum_{i=1}^na_i^b.
\end{equation}
For a given block length $n$, we generate the codebook of user $i=1,2$ by choosing $M_{i}$ sequences $\bx_i$ of length $n$ independently and uniformly over all the sequences of length $n$ and type $Q_i$ in $\mc{X}_i^n$. Note that $Q_i,i=1,2$ have rational entries with denominator $n$. We will write $\bx_{i,j}$ to denote the $j$-th codeword of user $i$. 

For a given channel output $\by_1\in \mc{Y}_1^n$, the best decoding rule to minimize the probability of error in decoding the message of user 1 is ML decoding, which consists of picking the message $m$ which maximizes $P(\by_1|\bx_{1,m})=\sum_{i=1}^{M_2}q_1^{(n)}(\by_1|\bx_{1,m},\bx_{2,i})/M_2$. Letting
\beq
q_{1,\calC_2}^{(n)}(\by_1|\bx_1)\dfn \frac{1}{M_2} \sum_{i=1}^{M_2} q_1^{(n)}(\by_1|\bx_1,\bx_{2,i})
\label{eqn:avgch}
\eeq
be the ``average'' channel observed at receiver 1, where the averaging is done over the codewords of user 2 in $\calC_2$, the decoding error probability at receiver 1 for transmitted codeword $\bx_{1,m}$ and codebooks $\calC_1$ and $\calC_2$ is given by:
\begin{align}
P_{e,1}(\bx_{1,m},\calC_1,\calC_2) =& \nonumber\\
\sum_{\by_1 \in \mc{Y}_1^n}P_{e,1}(&\bx_{1,m},\calC_1,\calC_2|\by_1) q_{1,\calC_2}^{(n)}(\by_1|\bx_{1,m})
\label{eqn:pe2}
\end{align}

With the introduction of the average channel (\ref{eqn:avgch}), and the use of two auxiliary parameters $(\rho,\lambda)\in [0,1]^2$, we can follow the approach of \cite{Gal68} to bound the conditional probability of decoding error $P_{e,1}(\bx_m,\calC_1,\calC_2|\by_1)$. Taking expectation over the random choice of codebooks $\calC_1$ and $\calC_2$  we obtain an average error probability:
\begin{align}
\overline{P}_{E_1}  
% \le& \bE_{\calC_2}\bE_{\calC_1}\bigg\{\sum_{\by_1 \in \mc{Y}_1^n}  [q_{1,\calC_2}^{(n)}(\by_1|\bX_{1,m})]^{\overline{\rho\lambda}}\nonumber\\
%&\bigg[\sum_{\substack{m'=1\\m'\ne m}}^{M_1} [q_{1,\calC_2}^{(n)}(\by_1|\bX_{1,m'})]^\lambda\bigg]^{\rho}\bigg\} \nonumber\\
\le& M_1^\rho\sum_{\by_1 \in \mc{Y}_1^n}\bE_{\calC_2}\bigg\{\bE_{\bX_1}\bigg[[q_{1,\calC_2}^{(n)}(\by_1|\bX_1)]^{\overline{\rho\lambda}}\bigg]\nonumber\\
& \cdot \bE_{\bX_1}^\rho\bigg[[q_{1,\calC_2}^{(n)}(\by_1|\bX_1)]^\lambda\bigg]\bigg\}
\label{eqn:pe3}
\end{align}
where we used Jensen's inequality to move the second expectation
inside $(\cdot)^\rho$.  

Equation (\ref{eqn:pe3}) is hard to handle, mainly due to the correlation introduced by $\calC_2$ between the two factors inside the outer expectation. Furthermore, the evaluation of the inner expectations over $\bX_1$ are complicated due to the powers $\overline{\rho\lambda}$ and $\lambda$ affecting $q_{1,\calC_2}^{(n)}(\by_1|\bX_1)$. Bounding methods based on Jensen's inequality and  
(\ref{distexp}) 
fail to give good results due to the loss of exponential tightness.

We proceed with a refined bounding technique based on the method of types inspired by \cite{Mer07}. While in this approach we still use (\ref{distexp}), we use it to bound sums with a number of terms that only grows polynomially with $n$, and as a result, exponential tightness is preserved.

Since the channel is memoryless,
\begin{align}
q_{1,\calC_2}^{(n)}(\by_1|\bx_1)=&\frac{1}{M_2}\sum_{i=1}^{M_2}\prod_{t=1}^n q_1(y_1(t)|x_{1}(t),x_{2,i}(t))\nonumber\\
=& \frac{1}{M_2} \sum_{\tp} N_{\bx_1,\by_1}(\tp)\nonumber\\
&\cdot e^{n \elq}
\label{eqn:qnc2}
\end{align}
where we used $N_{\bx_1,\by_1}(\tp)$ to denote the number of codewords $\bx_2$ in $\calC_2$ such that $(\bx_{1},\bx_2,\by_1)$ have empirical distribution $\tp$. We also used $\bE_{\hat{X}_1\hat{X}_2\hat{Y}_1}(\cdot)$ to denote expectation with respect to the distribution $\tp$.

Replacing (\ref{eqn:qnc2}) in (\ref{eqn:pe3}) and using (\ref{distexp}) 
three times, we obtain:
\begin{align}
\overline{P}_{E_1}\le&\frac{M_1^\rho}{M_2}\sum_{\hat{P}}\sum_{\hat{P}'}\sum_{\by_1 \in \mc{Y}_1^n} \bE_{\calC_2}\Bigg\{\bE_{\bX_1}\bigg[N_{\bX_1,\by_1}^{\overline{\rho\lambda}}(\hat{P})\bigg]\nonumber\\
&\cdot \bE_{\bX_1}^\rho\bigg[N_{\bX_1,\by_1}^{\lambda}(\hat{P}')\bigg]\Bigg\}\nonumber\\
&\cdot e^{n [\overline{\rho\lambda}E_{\hat{P}}\log q_1(\hat{Y}_1|\hat{X}_1,\hat{X}_2)+\lambda E_{\hat{P}'}\log q_1(\hat{Y}_1'|\hat{X}_1',\hat{X}_2')}
\label{eqn:pe4}
\end{align}
where we used $\hat{P}=\tp$ and $\hat{P}'=\tpp$ to shorten the expression.

We next consider the bounding of
\begin{align}
E(\by_1,\hat{P},\hat{P}')\dfn& \nonumber\\
 \bE_{\calC_2}\Bigg\{\bE_{\bX_1}&\bigg[N_{\bX_1,\by_1}^{\overline{\rho\lambda}}(\hat{P})\bigg]\bE_{\bX_1}^\rho\bigg[N_{\bX_1,\by_1}^{\lambda}(\hat{P}')\bigg]\Bigg\},
 \label{eqn:ec2}
\end{align}
and note that $N_{\bX_1,\by_1}(\hat{P})$ and $N_{\bX_1,\by_1}(\hat{P}')$ are formed by sums of an  exponentially large number of indicator functions, each of which takes value 1 with exponentially small probability. These sums concentrate around their means, which show different behavior depending on how the number of terms in the sum ($e^{n R_2}$) compares to the probability of each of the indicator functions taking value 1 (depending on the case considered, these probabilities take the form $e^{-n\Iyxz}$, $e^{-n\Iyxzp}$, or $e^{-n I(\hat{X}_2';\hat{Y}_1')}$). Whenever one of the factors in (\ref{eqn:ec2}) concentrates around its mean it behaves as a constant, and hence is uncorrelated with the remaining factor. As a result, the correlation between the two factors of (\ref{eqn:ec2}), which complicates the analysis, can be circumvented. We give the details of this part of the derivation in Appendix \ref{sec:proof}, but note here that the resulting bound on $E(\by_1,\hat{P},\hat{P}')$ depends on 
$\by_1$ only through a factor $1(\by_1\in P_{\hat{Y}_1},P_{\hat{Y}_1'}; P_{\hat{X}_1}=P_{\hat{X}_1'}=Q_1; P_{\hat{X}_2}=P_{\hat{X}_2'}=Q_2)$. Therefore, the innermost sum in (\ref{eqn:pe4}) can be evaluated by counting the number of vectors  $\by_1\in \mc{Y}_1^n$ that have empirical types $P_{\hat{Y}_1}$ and $P_{\hat{Y}_1'}$. Note that this count can only be positive for $P_{\hat{Y}_1}=P_{\hat{Y}_1'}$. This count is approximately equal to $e^{n H(\hat{Y}_1)}$ to first order in the exponent. Furthermore, the sums over $\hat{P}$ and $\hat{P}'$ in (\ref{eqn:pe4}) have a number of terms that only grows polynomially with $n$. Therefore, to first order, the exponential growth rate of (\ref{eqn:pe4}) equals the maximum exponential growth rate of the argument of the outer two sums, where the maximization is performed over the distributions $\hat{P}$ and $\hat{P}'$ which are rational, with denominator $n$. We can further upper bound the probability of error by enlarging the optimization region, maximizing over {\em any} probability distributions $\hat{P},\hat{P}'$. 

\section{Convex Optimization Issues}
\label{sec:convopt}
In order to get a valid evaluation of $E_{R,1}(R_1, R_2, Q_1,$ $Q_2, \rho, \lambda)$,
for any given $Q_1,Q_2,\rho,\lambda$ satisfying the 
constraints of the outer maximization, we need to 
accurately solve the inner minimization problems. A brute 
force search may not give accurate enough results in reasonable time. 
As will be shown below, the first minimization problem in 
(\ref{eqn:erx}) is a convex problem, and as a result, it that 
can be solved efficiently. In addition, convexity allows to 
lower bound the objective function by its supporting hyperplane, 
which in turn, allows to get a reliable\footnote{In our 
implementation we solve the original convex optimization 
problem using the MATLAB function {\tt fmincon}.} lower 
bound through the solution of a linear program.

The second minimization problem is not convex due to the 
non--convex constraint $R_2\le I(\hat{X}_2;\hat{Y}_1)$. 
If we remove this constraint, it will be later shown that 
we obtain a convex problem that can be solved efficiently. 
There are two possible situations: 

The first situation occurs when the optimal solution to the modified problem 
satisfies $R_2\le I(\hat{X}_2;\hat{Y}_1)$: in this case, 
the solution to the modified problem is also a solution to the 
original problem. 

The second situation is when the optimal solution to the modified problem 
satisfies $R_2>I(\hat{X}_2;\hat{Y}_1)$: 
in this case, a solution to the original problem must satisfy 
$R_2=I(\hat{X}_2;\hat{Y}_1)$. We prove this statement by contradiction. 
Let $P^*_1$ be the optimal solution to the modified problem, and $P^*_2$ be 
an optimal solution to the original problem. 
Now assume conversely, that there is no $P^*_2$ 
that satisfies $R_2=I(\hat{X}_2;\hat{Y}_1)$. With this assumption, we have 
that at $P^*_2$, $R_2<I(\hat{X}_2;\hat{Y}_1)$. Let 
$\mathcal{D}\triangleq \{P=(\tp,\tpp): P_{\hat{X}_1}=P_{\hat{X}_1'}=Q_1, 
P_{\hat{X}_2}=P_{\hat{X}_2'}=Q_2\}$. Note that $\mathcal{D}$ is a convex set 
and $P_1^*, P_2^* \in \mathcal{D}$. Due to the continuity of 
$I(\hat{X}_2;\hat{Y}_1)$, the 
straight line in $\mathcal{D}$ that joins $P^*_1$ and $P^*_2$ 
must pass through an intermediate point $\overline{P}=\alpha P^*_1 + 
(1-\alpha) P^*_2$, $\alpha \in (0,1)$, that satisfies 
$I(\hat{X}_2;\hat{Y}_1)=R_2$. Let $f_2(\cdot)$ be the objective function 
of the second minimization problem in (\ref{eqn:erx}), restricted to 
$\mathcal{D}$. It will be shown later that $f_2(\cdot)$, restricted to this 
domain, is a convex function. By hypothesis, $f_2(\overline{P})>f_2(P_2^*)$ and 
we have $f_2(P_1^*) \le f_2(P_2^*) < f_2(\overline{P})$. On the other hand, 
from the convexity of $f_2(\cdot)$, restricted to $\mathcal{D}$, we have 
$f_2(\overline{P}) \le \alpha f_2(P_1^*)+(1-\alpha) f_2(P_2^*)\le f_2(P_2^*)$ 
and we get a contradiction. Therefore, it follows that there is a solution $P^*_2$ to 
the original problem that satisfies $R_2=I(\hat{X}_2;\hat{Y}_1)$.

Let $f_1(\cdot)$ be the objective function of the first minimization 
problem in (\ref{eqn:erx}). First, we note that $P_2^*$ satisfies the 
constraints of the first minimization problem since they are less restrictive 
than the constraints of the second minimization problem in 
(\ref{eqn:erx}). We next prove that $f_1(P_2^*)= f_2(P_2^*)$. As a result, the 
optimal solution $P^*$ of the first minimization problem satisfies 
$f_1(P^*)\le f_1(P_2^*) = f_2(P^*_2)$, and we do not need to know 
$f_2(P_2^*)$ to evaluate the argument of the maximization in 
(\ref{eqn:erx}). Using the fact that at $P_2^*$, 
$I(\hat{X}_2;\hat{Y}_1)=I(\hat{X}_2';\hat{Y}_1')=R_2$, we have:
\begin{align}
&f_2(P_2^*)-f_1(P_2^*)\nonumber\\
&=\rho I(\hat{X}_1';\hat{X}_2',\hat{Y}_1') 
-\rho I(\hat{X}_1';\hat{Y}_1')-\rho (\Iyxzp-R_2) \nonumber\\
&=\rho \left[\Iyxzp-I(\hat{X}_2';\hat{Y}_1')-\Iyxzp+R_2\right]\nonumber \\
&=0,
\end{align}
where we used the identity 
$I(\hat{X}_1';\hat{X}_2',\hat{Y}_1')-I(\hat{X}_1';\hat{Y}_1')
=\Iyxzp-I(\hat{X}_2';\hat{Y}_1')$ in the second equality. 

In summary, if the solution to the second minimization problem in 
(\ref{eqn:erx}), without the constraint on $R_2$, satisfies 
$R_2>I(\hat{X}_2;\hat{Y}_1)$, then the first minimization problem in 
(\ref{eqn:erx}) dominates the expression. Otherwise, the solution to the 
second minimization problem in (\ref{eqn:erx}) without the constraint 
$R_2\le I(\hat{X}_2;\hat{Y}_1)$, equals the solution to the second 
minimization problem with this constraint.

It remains to show that the objective functions of the minimization problems 
in (\ref{eqn:erx}), $f_1(\tp,\tpp)$, $f_2(\tp,\tpp)$, restricted to the 
domain $\mathcal{D}$, are convex functions. 
Since the sum of convex functions is convex, to prove the convexity of 
$f_1(\cdot)$ on $\mathcal{D}$, we only need to prove that the different 
terms of
\begin{align}
f_1 = & -\overline{\rho\lambda}\bE_{\hat{X}_1\hat{X}_2\hat{Y}_1}\log 
q(\hat{Y}_1|\hat{X}_1,\hat{X}_2)-\nonumber\\
&\rho\lambda\bE_{\hat{X}_1'\hat{X}_2'\hat{Y}_1'}\log 
q(\hat{Y}_1'|\hat{X}_1',\hat{X}_2')
- H(\hat{Y}_1|\hat{X}_1)+\rho I(\hat{X}_1';\hat{Y}_1') \nonumber\\
&+ \max\bigg\{ \Iyxz-R_2; \nonumber\\
&\quad \quad \quad \quad \quad \quad \quad \quad \overline{\rho\lambda}(\Iyxz-R_2)\bigg\}\nonumber\\
&+ \max\bigg\{\overline{\rho}I(\hat{X}_2';\hat{Y}_1') +\rho \Iyxzp-R_2;\nonumber\\
&\rho (\Iyxzp-R_2); \rho \lambda(\Iyxzp-R_2)\bigg\} \label{eqn:f1a}
\end{align}
are convex within $\mathcal{D}$.

First, we have that $-\overline{\rho \lambda}\bE_{\hat{X}_1\hat{X}_2
\hat{Y}_1}\log q(\hat{Y}_1|\hat{X}_1,\hat{X}_2)-\rho\lambda\bE_{\hat{X}_1'
\hat{X}_2'\hat{Y}_1'}\log q(\hat{Y}_1'|\hat{X}_1',\hat{X}_2')$ 
is linear in $(\tp, \tpp)$ and therefore convex. Also, we have that 
$-H(\hat{Y}_1|\hat{X}_1)=H(\hat{X}_1)-H(\hat{X}_1,\hat{Y}_1)$ 
is convex for fixed $P_{\hat{X}_1}$ due to the concavity of 
$H(\hat{X}_1,\hat{Y}_1)$. 

In addition, $I(\hat{X}_1';\hat{Y}_1')$ can be written as 
$D(P_{\hat{X}_1'\hat{Y}_1'}||P_{\hat{X}_1'}\times 
P_{\hat{Y}_1'})$. Let $\overline{P}=\lambda \hat{P}+(1-\lambda) \check{P}$ for 
any $\hat{P}$, $\check{P}$ such that $\hat{P}_{\hat{X}_1'}=\check{P}_{\hat{X}_1'}$ 
and $\lambda \in [0,1]$. We have that $\overline{P}_{\hat{X}_1'\hat{Y}_1'}=
\lambda \hat{P}_{\hat{X}_1'\hat{Y}_1'}+(1-\lambda) 
\check{P}_{\hat{X}_1'\hat{Y}_1'}$ and $\overline{P}_{\hat{X}_1'}\times
\overline{P}_{\hat{Y}_1'}=\hat{P}_{\hat{X}_1'}\times(\lambda \hat{P}_{\hat{Y}_1'}+
(1-\lambda)\check{P}_{\hat{Y}_1'})=\lambda 
(\hat{P}_{\hat{X}_1'}\times\hat{P}_{\hat{Y}_1'})+(1-\lambda) 
(\check{P}_{\hat{X}_1'}\times\check{P}_{\hat{Y}_1'})$. The convexity of 
$\rho I(\hat{X}_1';\hat{Y}_1')$ for fixed $P_{\hat{X}_1'}$ follows from the 
convexity of $D(P\|Q)$ in the pair $(P,Q)$:
\begin{align}
I(\hat{X}_1';\hat{Y}'_1)\bigg|_{\overline{P}}&=
D(\overline{P}_{\hat{X}_1'\hat{Y}_1'}\|\overline{P}_{\hat{X}_1'}\times
\overline{P}_{\hat{Y}_1'})\nonumber\\
&\le \lambda D(\hat{P}_{\hat{X}_1'\hat{Y}_1'}\|\hat{P}_{\hat{X}_1'}\times
\hat{P}_{\hat{Y}_1'})\nonumber\\
&+(1-\lambda)D(\check{P}_{\hat{X}_1'\hat{Y}_1'}\|
\check{P}_{\hat{X}_1'}\times\check{P}_{\hat{Y}_1'})\nonumber\\
&= \lambda I(\hat{X}_1';\hat{Y}_1')\bigg|_{\hat{P}}+
(1-\lambda)I(\hat{X}_1';\hat{Y}_1')\bigg|_{\check{P}}.\nonumber\\
\label{eqn:convexI}
\end{align}

Continuing with the next term of (\ref{eqn:f1a}), 
\[
\max\big\{\Iyxz-R_2; \overline{\rho\lambda}(\Iyxz-R_2)\big\}
\]
we note that it is the maximum of two convex functions, and therefore convex. 
The convexity of each of the individual functions follows from the convexity of 
$\Iyxz$ for fixed $P_{\hat{X}_1}$, $P_{\hat{X}_2}$, which can be proved along 
the same lines as (\ref{eqn:convexI}).

Finally, we consider the last term of (\ref{eqn:f1a}):
\begin{align}
&\max\bigg\{\overline{\rho}I(\hat{X}_2';\hat{Y}_1')+\rho \Iyxzp-R_2;\nonumber\\
&\quad\quad\rho (\Iyxzp-R_2);  \rho \lambda(\Iyxzp-R_2) \bigg\}. \nonumber
\end{align}
Each of the arguments of the $\max\{\ldots \}$ can be shown to be the sum 
of convex functions for fixed $P_{\hat{X}_1'}$ and $P_{\hat{X}_2'}$, using a 
similar argument as the one used to prove (\ref{eqn:convexI}). Since the 
maximum of convex functions is convex, the convexity of $f_1$ restricted to 
$\mathcal{D}$ follows.

Using similar arguments, it is easy to show that 
\begin{align}
f_2&=-\overline{\rho\lambda}\bE_{\hat{X}_1\hat{X}_2\hat{Y}_1}
\log q_1(\hat{Y}_1|\hat{X}_1,\hat{X}_2)-\nonumber\\
&\rho\lambda\bE_{\hat{X}_1'\hat{X}_2'\hat{Y}_1'}\log 
q_1(\hat{Y}_1'|\hat{X}_1'\hat{X}_2')-
H(\hat{Y}_1|\hat{X}_1)+\nonumber\\ 
&\rho I(\hat{X}_1';\hat{X}_2,\hat{Y}_1')
 +\Iyxz -R_2 \nonumber
\end{align}
is convex in $\mathcal{D}$.

\section{Numerical Results}
\label{sec:results}
In this section, we present a numerical example to show the performance of the error exponent region introduced in Theorem \ref{thm:1}. We use as a baseline for comparison the error exponent region of Section~\ref{sec:background}, which is obtained with minor modifications from known results for single user and multiple access channels. 

We present results for the binary Z-channel model:
$Y_1=X_1*X_2\oplus Z$, $Y_2=X_2$, where $X_1,X_2,Y_1,Y_2 \in \{0,1\}$, $Z \sim \text{Bernoulli}(p)$, $*$ is multiplication, and $\oplus$ is modulo 2 addition. This is a modified version of the binary erasure IFC  studied in \cite{EO07}, where we add noise $Z$ to the received signal of user 1. In the results presented here, we fix $p=0.01$.

The boundary of the error exponent region is a surface in four dimensions $R_1,R_2,E_{R,1},E_{R,2}$.  This surface can be obtained parametrically by computing $E_{R,1},E_{R,2}$ as a function of $R_1, R_2, Q_1, Q_2$, by optimizing over $\rho$ and $\lambda$ in (\ref{eqn:erx}) and in the corresponding expression for $E_{R,2}$. The parameterization of $E_{R,i}$ in terms of $R_1, R_2, Q_1, Q_2$, allows the study of the error performance as a function of the parameters that directly influence it. 

\begin{figure}[htb]
\centerline{ \psfig{figure=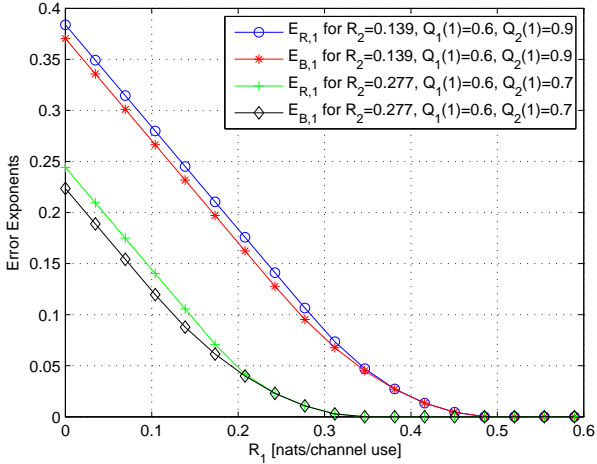,,width=3.5in}}
\vspace{-0.1in}
\caption{\footnotesize Error exponents as a function of $R_1$ for two different values of $R_2$ and fixed choices $Q_1, Q_2$. All the rates are in nats.}
\label{fig:plot1} 
%\vspace{-.1in}
\end{figure}
\vspace{0.05in}
\begin{figure}[htb]
\centerline{ \psfig{figure=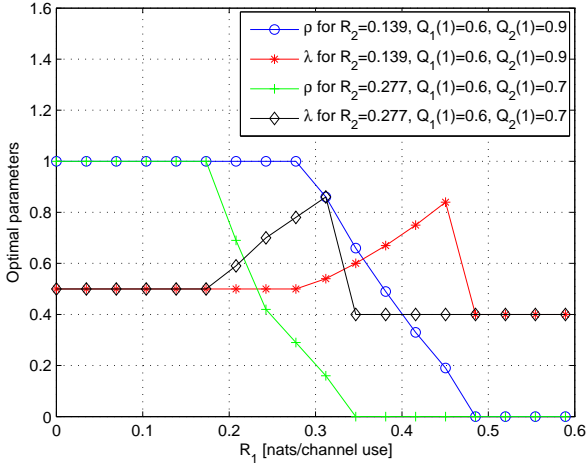,width=3.5in}}
\vspace{-0.1in}
\caption{\footnotesize Optimal parameters $\rho$ and $\lambda$ for the $E_{R,1}$ curves of Fig.  \ref{fig:plot1}. All the rates are in nats.}
\label{fig:plot2} 
%\vspace{-.1in}
\end{figure}

Fig. \ref{fig:plot1} shows that the error exponents under optimal decoding derived in this paper can be strictly better than the baseline error exponents of Section \ref{sec:background}. This suggests that the inequality obtained in Appendix \ref{sec:altexpr} for $R_1=0$ can be strict.  In addition, in all the plots that we computed for the Z-channel for different values of $Q_1, Q_2$ and $R_2$ we were not able to find a single case where the baseline exponent $E_{B,1}$ was larger than $E_{R,1}$. 

We see that the curves of $E_{R,1}$ ($E_{B,1}$) for fixed $R_2, Q_1, Q_2$ have a linear part for $R_1$ below a critical value $R_{1c}^{(R)}$ ($R_{1c}^{(B)}$), and a curvy part for $R_1>R_{1c}^{(R)}$ ($R_1>R_{1c}^{(B)}$) (note that the critical values depend on the parameters $R_2, Q_1$ and $Q_2$). Figure \ref{fig:plot2} shows the optimal parameters $\rho$ and $\lambda$ for the  $E_{R,1}$ curves shown in Fig. \ref{fig:plot1} for $R_2=0.139$ and $R_2=0.277$ nats/channel use. We see that for the linear part of the $E_{R,1}$ curves $\rho=1$ and $\lambda=1/2$ are optimal, while for the curvy part (i.e. $R_1>R_{1c}^{(R)}$) the optimal $\rho$ decreases to 0 and the optimal $\lambda$ increases towards 1. For $R_1$ in the interval $(0,\min\{R_{1c}^{(R)};R_{1c}^{(B)}\})$ the gap between the  $E_{R,1}$ and $E_{B,1}$ curves remains constant as both curves are lines with slope $-1$, and this gap is equal to the gap at $R_1=0$. In general, any gap between $E_{R,1}$ and $E_{G,1}$ at $R_1=0$ will remain constant in the interval where both curves have slope $-1$. We also note since the optimal parameters $\rho$ and $\lambda$ vary for different rates, these parameters are indeed active, i.e. they have influence on the resulting error exponent. 

\begin{figure}[htb]
\centerline{ \psfig{figure=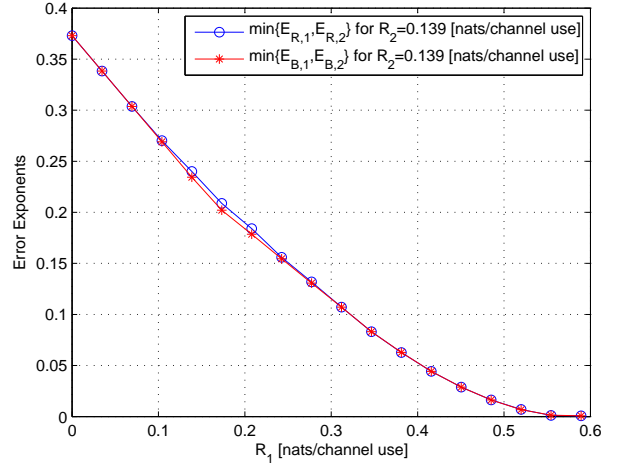,width=3.5in}}
\vspace{-0.1in}
\caption{\footnotesize Maximum error exponent simultaneously achievable for both users for fixed $R_2$ as a function of $R_1$. }
\label{fig:plot3} 
%\vspace{-.2in}
\end{figure}

The curves of Fig. \ref{fig:plot1} are obtained for fixed choices of $Q_1$ and $Q_2$, which are the distributions used to generate the random fixed composition codebooks. As $Q_1$ and $Q_2$ vary in the probability simplex $\calS$, we obtain the four-dimensional error exponent region $\{R_1, R_2, E_{R,1}(R_1,R_2,Q_1,Q_2), E_{R,2}(R_1,R_2,Q_1,Q_2): Q_1, Q_2 \in \calS\}$. In order to obtain a two-dimensional plot of the region, we consider a projection: we fix $R_2$ varying $R_1$ and plot the maximum value over $ Q_1 $ and $ Q_2 $ in the error exponent region of $\min\{E_{R,1}, E_{R,2}\}$. This corresponds to choosing $Q_1$ and $Q_2$ in order to maximize the error exponent {\em simultaneously} achievable for both users. Figure \ref{fig:plot3} shows this projection for $R_2=0.139$ and $R_2=0.277$ nats/channel use, where, for reference, we included the corresponding curves for the error exponents $E_{B,1}, E_{B,2}$ of Section \ref{sec:background}.

For the noiseless binary channel of user 2, $E_{R,2}=\max\{H(Q_2)-R_2;0\}$, and as a result, $E_{R,2}$ decreases with increasing $\pr(X_2=1)$ for $\pr(X_2=1)\ge 1/2$. On the other hand, because of the multiplication between $X_1$ and $X_2$ in the received signal $Y_1$, increasing $\pr(X_2=1)$ results in less interference for user 1, and a larger value of $E_{R,1}$. It follows that there is a direct trade-off between $E_{R,1}$ and $E_{R,2}$ through the choice of $Q_2$, and whenever $\min\{E_{R,1}, E_{R,2}\}$ is maximized, $E_{R,1}= E_{R,2}$. Therefore, in the curve of Fig. \ref{fig:plot3}, $E_{R,1}= E_{R,2}$.

From the plots of Figs. \ref{fig:plot1} and \ref{fig:plot3}, we see that the error exponents obtained from Theorem \ref{thm:1} sometimes outperform and are never worse than the baseline error exponents of Section~\ref{sec:background}.

\appendices
\section{Proof of Theorem \ref{thm:1}}
\label{sec:proof}
\renewcommand{\theequation}{\thesection.\arabic{equation}}
\setcounter{equation}{0}

\begin{figure}[htb]
\centerline{ \psfig{figure=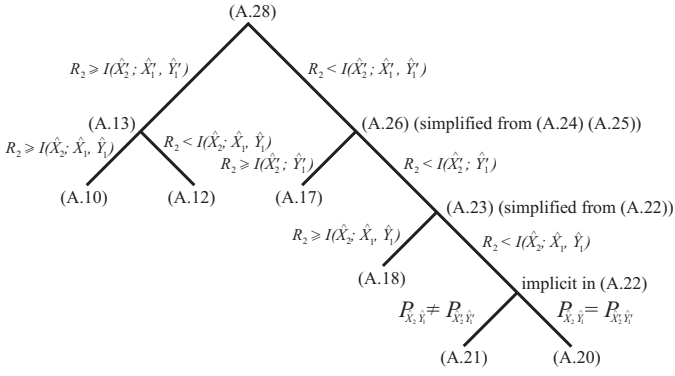,,width=3.5in}}
\vspace{-0.1in}
\caption{\footnotesize Tree representing the multiple ranges of $R_2$ considered in the derivation, and the equations that consolidate the different ranges.}
\label{fig:consolidation} 
\end{figure}

It is easy to see that the optimum decoder for user 1
picks the message $m$ ($1\le m\le M_1$) that 
maximizes $(1/M_2)\sum_{\bx_2\in\calC_2}q_1^{(n)}(\by_1|\bx_1,\bx_2)$, where
$M_1=\lceil e^{nR_1} \rceil$ and $M_2=\lceil e^{nR_2} \rceil$.
Applying Gallager's general upper bound to the ``channel''
$P(\by_1|\bx_1)=\frac{1}{M_2}\sum_{\bx_2\in\calC_2}q_1^{(n)}
(\by_1|\bx_1,\bx_2)$, 
we have for user no.\ 1:
\begin{align}
P_{E_1}&\le\sum_{\by_1}\left[\frac{1}{M_2}\sum_{\bx_2
\in\calC_2}
q_1^{(n)}(\by_1|\bx_1,\bx_2)\right]^{\overline{\rho\lambda}}\times\nonumber\\
&\left[\sum_{\bx_1'\ne \bx_1}\left(\frac{1}{M_2}\sum_{\bx_2\in\calC_2}
q_1^{(n)}(\by_1|\bx_1',\bx_2)\right)^\lambda
\right]^\rho,
\end{align}
where $\lambda\ge 0$ and $\rho\ge 0$ are arbitrary parameters to be
optimized in the sequel.
Thus, the average error probability 
is upper bounded by the expectation of the above
w.r.t.\ the ensemble of codes of both users. 
Let us take the expectation
w.r.t.\ the ensemble of user 1 first, and we denote this expectation
operator by $\bE_{\calC_1}\{\cdot\}$. Since the codewords of user 1 are
independent, the expectation of the summand in the sum above
is given by the product of expectations, namely, the product of
\begin{align}
A&\eqd\bE_{\calC_1}\left\{\left[\frac{1}{M_2}\sum_{\bx_2\in\calC_2}
q_1^{(n)}(\by_1|\bx_1,\bx_2)\right]^{\overline{\rho\lambda}}\right\}\nonumber\\
&=M_2^{\rho\lambda-1}
\bE_{\calC_1}\left\{\left[\sum_{\bx_2
\in\calC_2}
q_1^{(n)}(\by_1|\bx_1,\bx_2)\right]^{\overline{\rho\lambda}}\right\}.
\end{align}
and
\begin{align}
B&\eqd\bE_{\calC_1}\left\{\left[\sum_{\bx_1'\ne \bx_1}\left(\frac{1}{M_2}
\sum_{\bx_2\in\calC_2}
q_1^{(n)}(\by_1|\bx_1',\bx_2)\right)^\lambda\right]^\rho\right\}\nonumber\\
&=M_2^{-\rho\lambda}\bE_{\calC_1}\left\{\left[\sum_{\bx_1'\ne\bx_1}
\left(\sum_{\bx_2\in\calC_2}
q_1^{(n)}(\by_1|\bx_1',\bx_2)\right)^\lambda\right]^\rho\right\}.\nonumber
\end{align}
Now, let
$N_{\bx_1,\by_1}
(P_{\hat{X}_1\hat{X}_2\hat{Y}_1})$ 
denote the number of codewords $\{\bx_2\}$
that form a joint empirical PMF 
$P_{\hat{X}_1\hat{X}_2\hat{Y}_1}$ 
together with a given $\bx_1$ and $\by_1$.
Then, using (\ref{distexp}), $A$ can be bounded by
\begin{align}
A=&M_2^{\rho\lambda-1} 
\bE_{\bX_1}\Bigg[\sum_{P_{\hat{X}_1\hat{X}_2\hat{Y}_1}} 
N_{\bX_1,\by_1}
(P_{\hat{X}_1\hat{X}_2\hat{Y}_1})\times\nonumber\\
&\quad\quad\quad\quad\quad\quad\quad
e^{n\bE_{\hat{X}_1\hat{X}_2\hat{Y}_1}
  \log q_1(\hat{Y}_1|\hat{X}_1,\hat{X}_2)}\Bigg]^{\overline{\rho\lambda}}\nonumber\\
\le&
M_2^{\rho\lambda-1} 
\sum_{P_{\hat{X}_1\hat{X}_2\hat{Y}_1}} 
\bE_{\bX_1}N_{\bX_1,\by_1}^{\overline{\rho\lambda}}
(P_{\hat{X}_1\hat{X}_2\hat{Y}_1})\times\nonumber\\
&\quad\quad\quad\quad\quad\quad\quad
e^{n\overline{\rho\lambda}\bE_{\hat{X}_1\hat{X}_2\hat{Y}_1}
  \log q_1(\hat{Y}_1|\hat{X}_1,\hat{X}_2)}
\end{align}
where 
$q_1(\hat{Y}_1|\hat{X}_1,\hat{X}_2)$ is the 
single--letter transition probability distribution
of the IFC, and
where $\bE_{\hat{X}_1\hat{X}_2\hat{Y}_1}f(\hat{X}_1,\hat{X}_2,\hat{Y}_1)$, 
for a generic function $f$,
denotes the expectation operator when the 
RV's $(\hat{X}_1,\hat{X}_2,\hat{Y}_1)$ are understood
to be distributed according to 
$P_{\hat{X}_1\hat{X}_2\hat{Y}_1}$. Similarly,
(and using Jensen's inequality to push the expectation 
w.r.t.\ $ \calC_1 $ into the brackets), we have:
\begin{align}
B\leq& M_2^{-\rho\lambda}M_1^{\rho}\Bigg[
  \sum_{P_{\hat{X}_1\hat{X}_2\hat{Y}_1}} \bE_{\bX_1} 
N_{\bX_1,\by_1}^{\lambda}
(P_{\hat{X}_1\hat{X}_2\hat{Y}_1})\times
\nonumber\\
&\quad\quad\quad\quad\quad\quad\quad
 e^{n\lambda
  \bE_{\hat{X}_1\hat{X}_2\hat{Y}_1}\log 
q(\hat{Y}_1|\hat{X}_1,\hat{X}_2)}\Bigg]^\rho
\end{align}
Taking the product of these two expressions, applying (\ref{distexp})
to the summation in the bound for $ B $, and taking
expectations with respect to the codebook $\calC_2$ yields
\begin{align}
\label{eqn:EcyAB}
\bE_{\calC_2}(A&B)\leq M_1^\rho M_2^{-1}
  \sum_{P_{\hat{X}_1\hat{X}_2\hat{Y}_1}}
  \sum_{P_{\hat{X}_1'\hat{X}_2'\hat{Y}_1'}}\nonumber\\
&\bE_{\calC_2}[\bE_{\bX_1}
N_{\bX_1,\by_1}^{\overline{\rho\lambda}}
(P_{\hat{X}_1\hat{X}_2\hat{Y}_1})
\bE^{\rho}_{\bX_1}
  N_{\bX_1,\by_1}^{\lambda}
(P_{\hat{X}_1'\hat{X}_2'\hat{Y}_1'})]\nonumber\\
&\times\exp\{n[\overline{\rho\lambda}\bE_{\hat{X}_1\hat{X}_2\hat{Y}_1}
  \log q_1(\hat{Y}_1|\hat{X}_1,\hat{X}_2)\nonumber\\
&\quad\quad\quad+\rho\lambda \bE_{\hat{X}_1'\hat{X}_2'\hat{Y}_1'}
  \log q_1(\hat{Y}_1'|\hat{X}_1',\hat{X}_2')]\}
\end{align}
The next step is to bound the term involving the expectation over
$\calC_2$. 
As noted, the codewords $\{\bX_1\}$ and $
\{\bX_2\} $ are 
randomly selected i.i.d.\ over the type classes $\calT_1=\calT_{Q_1}$
and $\calT_2=\calT_{Q_2}$
corresponding to probability distributions $Q_1$ and $Q_2$, respectively.
To avoid cumbersome notation, we denote hereafter
$\hat{P}=\tp$ and
$\hat{P}'=\tpp$
and assume that $P_{\hat{X}_1}=P_{\hat{X}_1'}=Q_1$, $P_{\hat{X}_2}=P_{\hat{X}_2'}=Q_2$,
$P_{\hat{Y}_1}=P_{\hat{Y}_1'}$ and that $\by_1$ lies in the type class
corresponding to $P_{\hat{Y}_1}$.
We will also use the shorthand notation
\beq
\bE_{\calC_2}\triangleq \bE_{\calC_2}[\bE_{\bX_1}
N_{\bX_1,\by_1}^{\overline{\rho\lambda}}(\hat{P})
\bE_{\bX_1}^\rho N_{\bX_1,\by_1}^{\lambda}
(\hat{P}')].
\eeq
The bounding of $\bE_{\calC_2}$ requires considering multiple cases which depend on how $R_2$ compares to different information quantities, and also depend on properties of the joint types $\tp, \tpp$. In order to guide the reader through the different steps we present in Fig. \ref{fig:consolidation} below a schematic representation of the different cases that arise.

We first consider two different ranges of $R_2$,
according to its comparison with $\Iyxzp$:

\vspace{0.5cm}

\noindent
{\bf 1. The range $R_2 \ge \Iyxzp$.} Here we have:
\begin{align}
&\bE_{\calC_2}=\bE_{\calC_2}
\Bigg\{\bE_{\bX_1}\left[N_{\bX_1,\by_1}^{1-\rho \lambda}(\ph)\right]\nonumber\\
&\quad\quad\quad\quad\quad\quad\quad\quad\quad\quad
\times\bigg[\frac{1}{|\calT_1|}\sum_{\tilde{\bx}\in \calT_1}
N_{\tilde{\bx}_1,\by_1}^\lambda(\php)\bigg]^\rho \Bigg\}\nonumber\\
=&\bE_{\calC_2}\Bigg\{\bE_{\bX_1}\left[ 
N_{\bX_1,\by_1}^{1-\rho \lambda}(\ph)\right]\cdot
\bigg[\frac{1}{|\calT_1|}\sum_{\tilde{\bx} \in \calT_1}
N_{\tilde{\bx}_1,\by_1}^\lambda(\php)\bigg]^\rho\times\nonumber\\
&1\left[N_{\tilde{\bx}_1,\by_1}(\php)\le e^{n[(R_2-\Iyxzp)+\epsilon]}, 
\forall \tilde{\bx}_1\in \calT_1\right]\Bigg\}\nonumber\\
&+\bE_{\calC_2}\Bigg\{\bE_{\bX_1}\left[ 
N_{\bX_1,\by_1}^{1-\rho \lambda}(\ph)\right]\bigg[\frac{1}
{|\calT_1|}\sum_{\tilde{\bx} \in \calT_1}
N_{\tilde{\bx_1},\by_1}^\lambda(\php)\bigg]^\rho\times\nonumber\\
&1\left[\exists \tilde{\bx}\in \calT_1: N_{\tilde{\bx}_1,\by_1}
(\php)> 
e^{n[(R_2-\Iyxzp)+\epsilon]} \right]\Bigg\}\nonumber\\
\le&\bE_{\calC_2}\Bigg\{\bE_{\bX_1}\left[ 
N_{\bX_1,\by_1}^{\overline{\rho\lambda}}(\ph)\right]\cdot
\bigg[e^{-n (H(\hat{X}_1')-\epsilon)}\times\nonumber\\
&\sum_{\tilde{\bx} \in \calT_1}
1\left[(\tilde{\bx},\by_1)\in \calT_{P_{\hat{X}_1'\hat{Y}_1'}}\right]
\cdot e^{n\lambda (R_2-\Iyxzp+\epsilon)}\bigg]^\rho \Bigg\}\nonumber\\
& + e^{n R_2}\mbox{Pr}\left[\exists \tilde{\bx}\in \calT_1: 
N_{\tilde{\bx},\by_1}(\php)> e^{n[(R_2-\Iyxzp)+\epsilon]} \right]\nonumber\\
\stackrel{.}{\le}&
\bE_{\calC_2}\left\{\bE_{\bX_1}\left[ 
N_{\bX_1,\by_1}^{1-\rho \lambda}(\ph)\right]\right\}\cdot
e^{-n \rho [H(\hat{X}_1')-H(\hat{X}_1'|\hat{Y}_1')]}\times
\nonumber\\
&\quad\quad\quad\quad\quad\quad\quad\quad\quad\quad\quad
 e^{n\rho \lambda (R_2-\Iyxzp)}
\end{align}
where in the second to last inequality we used $ N_{\bx_1,\by} \leq M_2 $, 
and in the last inequality we used the fact that 
\begin{align}
&\mbox{Pr}\left\{
\exists \tilde{\bx}\in \calT_1: 
N_{\tilde{\bx},\by_1}(\php)> e^{n[(R_2-\Iyxzp)+\epsilon]} \right\}\nonumber\\
&\le e^{n (H(\hat{X}_1')+\epsilon)}\cdot\mbox{Pr}\left\{
N_{\tilde{\bx},\by_1}(\php)> e^{n[(R_2-\Iyxzp)+\epsilon]} \right\}\nonumber\\
\end{align}
for any $\tilde{\bx} \in \calT_1$, 
which decays doubly exponentially with $n$
(cf.\ \cite[Appendix]{Mer07}).

To compute $\bE_{\calC_2}\left\{\bE_{\bX_1}
\left[ N_{\bX_1,\by_1}^{1-\rho \lambda}(\ph)\right]\right\}$ 
we consider two cases, according to the
comparison between $R_2$ and $\Iyxz$:

\vspace{0.25cm}

\noindent
{\it The case $R_2 \ge \Iyxz$}. Here, we have:
\begin{align}
&\bE_{\calC_2}\bE_{\bX_1}\left[
N_{\bX_1,\by_1}^{1-\rho \lambda}(\ph)\right] = 
\bE_{\bX_1}\bE_{\calC_2}\left[ 
N_{\bX_1,\by_1}^{1-\rho \lambda}(\ph)\right] \nonumber\\
&\stackrel{.}{\le}\bE_{\bX_1}\left[ 1\left((\bX_1,\by_1)\in 
\calT_{P_{\hat{X}_1\hat{Y}_1}}\right)e^{n\overline{\rho\lambda}
(R_2-\Iyxz)}\right] \nonumber\\
&\stackrel{.}{=}e^{-n I(\hat{X}_1;\hat{Y}_1)} 
e^{n\overline{\rho\lambda}(R_2-\Iyxz)}.
\label{eqn:EcyExNa}
\end{align}

% \begin{align}
% E_{C_Y}E_{\bX}\left[ N_{\bX,\by}^{1-\rho \lambda}(\ph)\right] &= E_{C_Y}\left[\frac{1}{|Q_X|}\sum_{{\bx} \in Q_X}N_{{\bx},\by}^{1-\rho \lambda}(\ph)\right]\nonumber\\
% &= E_{C_Y}\left\{\frac{1}{|Q_X|}\sum_{\bx \in Q_X} N_{\bx,\by}^{1-\rho \lambda}(\ph) 1\left[N_{\bx,\by}(\ph)\le e^{n[(R_y-\Iyxz)+\epsilon]}, \forall {\bx}\in Q_X \right]\right\}\nonumber\\
% &+E_{C_Y}\left\{\frac{1}{|Q_X|}\sum_{{\bx} \in Q_X}N_{{\bx},\by}^{\overline{\rho\lambda}}(\ph) 1\left[\exists {\bx}\in Q_X: N_{{\bx},\by}(\ph)> e^{n[(R_y-\Iyxz)+\epsilon]} \right]\right\}\nonumber\\
% &\le  E_{C_Y}\left\{e^{-n (H(\hat{X})-\epsilon)}\sum_{{\bx} \in Q_X}1\left[({\bx},\by)\in P_{\hat{X},\hat{Z}}\right]\cdot e^{n\overline{\rho\lambda} (R_y-\Iyxz+\epsilon)} \right\}\nonumber\\
% & + e^{n R_y} Pr\left[\exists {\bx}\in Q_X: N_{{\bx},\by}(\ph)> e^{n[(R_y-\Iyxz)+\epsilon]} \right]\nonumber\\
% &\stackrel{.}{\le} E_{C_Y}\left\{e^{-n [H(\hat{X})-H(\hat{X}|\hat{Z})]}\cdot e^{n\overline{\rho\lambda} (R_y-\Iyxz)}\right\}\nonumber\\
% &= e^{-n [H(\hat{X})-H(\hat{X}|\hat{Z})]}\cdot e^{n\overline{\rho\lambda} (R_y-\Iyxz)}
% \label{eqn:EcyExNa}
% \end{align}
% where in the last inequality we used the fact that $Pr\left[\exists {\bx}\in Q_X: N_{{\bx},\bz}(\ph)> e^{n[(R_y-\Iyxz)+\epsilon]} \right]$ decays doubly exponentially with $n$.

Therefore, when
$$R_2 \ge \max\{\Iyxz,\Iyxzp\}$$
we have:
\begin{align}
\label{eqn:Ecy1a}
\bE_{\calC_2}&\stackrel{.}{\le}
\exp\left\{n\left[-I(\hat{X}_1;\hat{Y}_1)+\overline{\rho\lambda}(R_2-\Iyxz) 
\right.\right.\nonumber\\
&\quad\left.\left.-\rho I(\hat{X}_1';\hat{Y}_1')+\rho\lambda(R_2-\Iyxzp)\right] \right\}.
\end{align}

\vspace{0.25cm}

\noindent
{\it The case $R_2 < \Iyxz$}. Here we have:
\begin{align}
\bE_{\calC_2}\bE_{\bX_1} 
\left[N_{\bX_1,\by_1}^{\overline{\rho\lambda}}(\ph)\right]&\le 
\bE_{\calC_2}\bE_{\bX_1}\left[ N_{\bX_1,\by_1}(\ph)\right]\nonumber\\
&\stackrel{.}{\le} e^{-n I(\hat{X}_1;\hat{Y}_1)}\cdot e^{n(R_2-\Iyxz)},
\label{eqn:EcyExNb}
\end{align}
where we used the fact that $\overline{\rho\lambda} \le 1$ and then
estimated the expectation of $ N_{\bX_1,\by_1}(\ph)$ as $M_2$ times
the probability $\bx_2$ would fall into the corresponding 
conditional type.
Therefore, when 
$$\Iyxzp \le R_2 < \Iyxz$$
we have:
\begin{align}
\label{eqn:Ecy1b}
\bE_{\calC_2}&\stackrel{.}{\le}
\exp\left\{n\left[-I(\hat{X}_1;\hat{Y}_1) 
+(R_2-\Iyxz)\right.\right.\nonumber\\
&\quad\left.\left.-\rho I(\hat{X}_1';\hat{Y}_1')+
\rho\lambda(R_2-\Iyxzp)\right] \right\}.
\end{align}

\vspace{0.5cm}
%%%%%%%%%%%%%%%%%%%%%%%%%%%%%%%%%%
%[SUMMARIZE ``RANGE 1'' USING MAX INSTEAD OF INDICATORS]
%%%%%%%%%%%%%%%%%%%%%%%%%%%%%%%%%%
The exponents for the subcases $ (\ref{eqn:Ecy1a}) $ and $
(\ref{eqn:Ecy1b}) $ corresponding 
to $ R_2 \geq \Iyxz $ and $ R_2 < \Iyxz $, respectively, differ only
in the factors ($ \overline{\rho\lambda} $ and $ 1 $, resp.) multiplying the term $
R_2 - \Iyxz $.  
Therefore, we can consolidate these two subscases of $
R_2 \geq \Iyxzp $ into the expression:
\begin{align}
\bE_{\calC_2}&\stackrel{.}{\le}
\exp\left\{n\left[-I(\hat{X}_1;\hat{Y}_1) 
+\right.\right.\nonumber\\
& \quad\min\{\overline{\rho\lambda}(R_2 - \Iyxz), \nonumber \\
&\quad\quad\quad (R_2-\Iyxz)\} \nonumber \\
&\quad\left.\left.-\rho I(\hat{X}_1';\hat{Y}_1')+
\rho\lambda(R_2-\Iyxzp)\right] \right\}, \label{eqn:Ecy1}
\end{align}
since $
\min\{\overline{\rho\lambda} $ $ (R_2 - \Iyxz), $ $ (R_2-\Iyxz)\} $ is $
\overline{\rho\lambda} $ $ (R_2 - \Iyxz) $ when $ R_2 \geq \Iyxz $ and $
(R_2-\Iyxz) $  when $ R_2 < \Iyxz $.

\noindent
{\bf 2. The range $R_2 < \Iyxzp$}.
In this range,
\begin{align}
\bE_{\calC_2}&=\bE_{\calC_2}
\left\{\bE_{\bX_1}\left[N_{\bX_1,\by_1}^{1-\rho \lambda}(\ph)\right]
\bE_{\bX_1}^\rho\left[N_{\bX_1,\by_1}^\lambda(\php)\right] \right\}\nonumber\\
&\le\bE_{\calC_2}\left\{\bE_{\bX_1}\left[
N_{\bX_1,\by_1}^{1-\rho \lambda}(\ph)\right]\bE_{\bX_1}^\rho
\left[N_{\bX_1,\by_1}(\php)\right]\right\}\nonumber\\
\end{align}
where we assumed $\lambda \le 1$ in the last step. 
The second expectation over $ \bX_1$ can be evaluated as
\begin{align}
E_{\bX_1}
N_{\bX_1,\by_1}&(P_{\hat{X}_1'\hat{X}_2'\hat{Y}_1'})\nonumber\\
&= \sum_{\bx_2\in\calC_2} \bE_{\bX_1}1((\bX_1,\bx_2,\by_1) \in
\calT_{P_{\hat{X}_1'\hat{X}_2'\hat{Y}_1'}})\nonumber\\
&\exe e^{-nI(\hat{X}_1';\hat{X}_2',\hat{Y}_1')} 
\sum_{\bx_2\in\calC_2} 1((\bx_2,\by_1) \in
\calT_{P_{\hat{X}_2'\hat{Y}_1'}})\nonumber\\
&=e^{-nI(\hat{X}_1';\hat{X}_2',\hat{Y}_1')} N_{\by_1}
(P_{\hat{X}_2'\hat{Y}_1'}),
\end{align}
where $N_{\by_1}(P_{\hat{X}_2'\hat{Y}_1'})$
is the number of codewords $\{\bx_2\}$ that are jointly typical with
$\by_1$ according to $P_{\hat{X}_2'\hat{Y}_1'}$.
Thus,
\begin{align}
&\bE_{\calC_2}\big[\bE_{\bX_1}
N_{\bX_1,\by_1}^{\overline{\rho\lambda}}(\hat{P})
\bE^{\rho}_{\bX_1}
N_{\bX_1,\by_1}(\hat{P}')\big]\nonumber\\
&\exe
e^{-n\rho I(\hat{X}_1';\hat{X}_2',\hat{Y}_1')} 
\bE_{\calC_2}\big[\bE_{\bX_1}
N_{\bX_1,\by_1}^{\overline{\rho\lambda}}(
P_{\hat{X}_1\hat{X}_2\hat{Y}_1})
N_{\by_1}^{\rho}(P_{\hat{X}_2'\hat{Y}_1'})\big] \nonumber\\
&=
e^{-n\rho I(\hat{X}_1';\hat{X}_2',\hat{Y}_1')} 
\bE_{\bX_1} \bE_{\calC_2}\big[
N_{\bX_1,\by_1}^{\overline{\rho\lambda}}
(P_{\hat{X}_1\hat{X}_2\hat{Y}_1})
N_{\by_1}^{\rho}(P_{\hat{X}_2'\hat{Y}_1'})\big].\label{eqn:Ecy2outer}
\end{align}
To bound $\bE_{\bX_1}\bE_{\calC_2}[
N_{\bX_1,\by_1}^{\overline{\rho\lambda}}(\hat{P})
N_{\by_1}^{\rho}(\hat{P}')]$, we consider two 
cases depending on how $R_2$ compares to $\Iyzp$. 

\vspace{0.25cm}

\noindent
{\it The case $R_2 \ge \Iyzp$}. Here, we have:
\begin{align}
&\bE_{\bX_1}\bE_{\calC_2}[N_{\bX_1,\by_1}^{\overline{\rho\lambda}}
(\hat{P})
N_{\by_1}^{\rho}(\hat{P}')]\nonumber\\
=&\bE_{\bX_1}\bE_{\calC_2}\Bigg\{N_{\bX_1,\by_1}^{\overline{\rho\lambda}}(\hat{P}) 
N_{\by_1}^{\rho}(\hat{P}')\times\nonumber\\
&\quad\quad\quad\quad\quad\quad\quad 1\bigg[N_{\by_1}(\hat{P}')\le e^{n(R_2-\Iyzp+\epsilon)}
\bigg]\Bigg\}\nonumber\\
&+\bE_{\bX_1}\bE_{\calC_2}\Bigg\{N_{\bX_1,\by_1}^{\overline{\rho\lambda}}(\hat{P}) 
N_{\by_1}^\rho(\hat{P}')\times\nonumber\\
& \quad\quad\quad\quad\quad\quad\quad 1\bigg[N_{\by_1}(\hat{P}')> 
e^{n(R_2-\Iyzp+\epsilon)}\bigg] \Bigg\}\nonumber\\
\stackrel{.}{\le}& e^{n\rho(R_2-\Iyzp)}\bE_{\bX_1}\bE_{\calC_2}
\bigg[N_{\bX_1,\by_1}^{\overline{\rho\lambda}}(\hat{P}) \bigg]\nonumber\\
&+ e^{n(\overline{\rho\lambda}+\rho)R_2}\mbox{Pr}
\bigg[N_{\by_1}(\hat{P}')>e^{n(R_2-\Iyzp+\epsilon)} \bigg] \nonumber\\
\stackrel{.}{\le}& \exp\Bigg\{n\Bigg[\rho (R_2-\Iyzp) - \Ixz \nonumber\\
&+1(R_2 \ge \Iyxz)\overline{\rho\lambda}(R_2-\Iyxz)\nonumber\\
&+1(R_2 < \Iyxz)(R_2-\Iyxz) \bigg] \Bigg\} \nonumber \\
 = & \exp\Bigg\{n\Bigg[\rho (R_2-\Iyzp) - \Ixz \nonumber\\
&+\min\{\overline{\rho\lambda}(R_2-\Iyxz),\nonumber\\
&\quad\quad(R_2-\Iyxz)\} \bigg] \Bigg\} \label{eqn:Ecy2a}
\end{align}
%%%%%%%%%%%%%%%%%%%%%%%%%%%%%%%%%%
%[CAN ADD A NEW LINE REPLACING INDICATORS WITH MAX]
%%%%%%%%%%%%%%%%%%%%%%%%%%%%%%%%%%
where we used the fact that 
$\mbox{Pr}\big[N_{\by_1}(\hat{P}')>e^{n(R_2-\Iyzp+\epsilon)} \big]$ 
decays doubly exponentially in the third inequality, 
and bounded $\bE_{\bX_1}\bE_{\calC_2}\big[N_{\bX_1,\by_1}^{\overline{\rho\lambda}}
(\hat{P}) \big]$ using (\ref{eqn:EcyExNa}) and (\ref{eqn:EcyExNb}) 
in the last inequality.

\vspace{0.25cm}

%%%%%%%%%%%%%%%%%%%%%%%%%%%%%%%%%%
%[CAN ADD A SUMMARY OF THIS RANGE/CASE]
%%%%%%%%%%%%%%%%%%%%%%%%%%%%%%%%%%

\noindent
{\it The case $R_2 < \Iyzp$}. Here,
we further split the evaluation into two parts.
In the first part, $R_2 \ge \Iyxz$, and we have:
\begin{align}
&\bE_{\bX_1}\bE_{\calC_2}
[N_{\bX_1,\by_1}^{\overline{\rho\lambda}}(\hat{P})
N_{\by_1}^{\rho}(\hat{P}')]\nonumber\\
\le& 
\bE_{\bX_1}\bE_{\calC_2}\Bigg\{N_{\bX_1,\by_1}^{1-\rho \lambda}
(\hat{P}) N_{\by_1}^{\rho}(\hat{P}')\times\nonumber\\
&\quad\quad\quad\quad\quad\quad 1\bigg[N_{\bX_1,\by_1}(\hat{P})\le 
e^{n(R_2-\Iyxz+\epsilon)}\bigg]\Bigg\}\nonumber\\
&+\bE_{\bX_1}\bE_{\calC_2}\Bigg\{N_{\bX_1,\by_1}^{1-\rho \lambda}(\hat{P}) 
N_{\by_1}^\rho(\hat{P}')\times\nonumber\\
&\quad\quad\quad\quad\quad\quad 1\bigg[N_{\bX_1,\by_1}(\hat{P})> 
e^{n(R_2-\Iyxz+\epsilon)}\bigg] \Bigg\}\nonumber\\
\stackrel{.}{\le}&e^{n\overline{\rho\lambda}(R_2-\Iyxz)}\times\nonumber\\
&\quad\quad\quad\quad\quad\bE_{\bX_1}\bE_{\calC_2} 
\bigg\{N_{\by_1}^{\rho}(\hat{P}')1\big[(\bX_1,\by_1)\in 
\calT_{P_{\hat{X}_1\hat{Y}_1}}\big] \bigg\}\nonumber\\
&+ e^{n(\overline{\rho\lambda}+\rho)R_2}\mbox{Pr}\bigg[N_{\bX_1,\by_1}
(\hat{P})>e^{n(R_2-\Iyxz+\epsilon)} \bigg] \nonumber\\
\stackrel{.}{\le}& e^{n[\overline{\rho\lambda}(R_2-\Iyxz)-\Ixz]}
\bE_{\calC_2}\big[N_{\by_1}^{\rho}(P_{\hat{X}_2'\hat{Y}_1'})\big]\nonumber\\
\stackrel{.}{\le}&\exp\bigg\{n\big[\overline{\rho\lambda}(R_2-\Iyxz)-
\Ixz \nonumber\\
& \quad\quad\quad\quad\quad\quad \quad\quad\quad\quad\quad\quad +
R_2-\Iyzp \big] \bigg\} \label{eqn:Ecy2ba}
\end{align}
where we used in the last inequality
\[
\bE_{\calC_2}\big[N_{\by_1}^{\rho}(P_{\hat{X}_2'\hat{Y}_1'})\big]
\le \bE_{\calC_2}\big[N_{\by_1}(P_{\hat{X}_2'\hat{Y}_1'})\big]
\stackrel{.}{=}  e^{n(R_2-\Iyzp)}
\]
valid for $\rho \le 1$.

%%%%%%%%%%%%%%%%%%%%%%%%%%%%%%%%%%
%[CAN ADD A SUMMARY OF THIS RANGE/CASE]
%%%%%%%%%%%%%%%%%%%%%%%%%%%%%%%%%%

The other part corresponds to $R_2 < \Iyxz$.
Here we have:
\begin{align}
\bE_{\bX_1}&\bE_{\calC_2}[
N_{\bX_1,\by_1}^{\overline{\rho\lambda}}(\hat{P})
N_{\by_1}^{\rho}(\hat{P}')]\nonumber\\ 
=&
\bE_{\bX_1}\bE_{\calC_2}\bigg\{N_{\bX_1,\by_1}^{1-\rho \lambda}(\hat{P}) 
N_{\by_1}^{\rho}(\hat{P}')
1\big[N_{\by_1}(\hat{P}')\le e^{n\epsilon}\big]\bigg\}\nonumber\\
&+\bE_{\bX_1}\bE_{\calC_2}\bigg\{N_{\bX_1,\by_1}^{\overline{\rho\lambda}}(\hat{P}) 
N_{\by_1}^\rho(\hat{P}')
1\big[N_{\by_1}(\hat{P}')> e^{n\epsilon}\big] \bigg\}\nonumber\\
\le& e^{n\rho\epsilon}\bE_{\bX_1}\bE_{\calC_2} 
\bigg\{N_{\bX_1,\by_1}^{\overline{\rho\lambda}}(\hat{P})
1\big[N_{\by_1}(\hat{P}') \ge 1 \big] \bigg\}\nonumber\\
&\quad\quad\quad\quad\quad\quad\quad
+ e^{n(\overline{\rho\lambda}+\rho)R_2}\mbox{Pr}
\bigg[N_{\by_1}(\hat{P}')>e^{n\epsilon} \bigg]\nonumber\\
\stackrel{.}{\le}& \bE_{\bX_1}\bE_{\calC_2}
\bigg\{N_{\bX_1,\by_1}^{\overline{\rho\lambda}}(\hat{P})\cdot
1\big[N_{\by_1}(\hat{P}') \ge 1 \big]\times\nonumber\\
& \quad\quad\quad\quad\quad\quad\quad\quad\quad\quad\quad\quad
1\big[N_{\bX_1,\by_1}
(\hat{P})\le e^{n\epsilon}\big]\bigg\}\nonumber\\
&+\bE_{\bX_1}\bE_{\calC_2}\bigg\{N_{\bX_1,\by_1}^{\overline{\rho\lambda}}
(\hat{P})\cdot 1\big[N_{\by_1}(\hat{P}') \ge 1 \big]\times\nonumber\\
&\quad\quad\quad\quad\quad\quad\quad\quad\quad\quad\quad\quad
1\big[N_{\bX_1,\by_1}(\hat{P})> e^{n\epsilon}\big]\bigg\}\nonumber\\
\stackrel{.}{\le}& e^{n\overline{\rho\lambda}\epsilon}
\bE_{\bX_1}\bE_{\calC_2}\bigg\{1\big[N_{\by_1}(\hat{P}') 
\ge 1 \big]\times\nonumber\\
& \quad\quad\quad\quad\quad\quad\quad\quad\quad\quad\quad\quad
1\big[N_{\bX_1,\by_1}
(\hat{P})\ge 1\big]\bigg\}\nonumber\\
&+ e^{n\overline{\rho\lambda}R_2}\bE_{\bX_1}\bigg\{\mbox{Pr} 
\big[N_{\bX_1,\by_1}(\hat{P})> e^{n\epsilon}\big]\bigg\}\nonumber\\
\stackrel{.}{=}& \frac{1}{|\calT_1|}\sum_{\tilde{\bx}_1\in \calT_1} 
1\big[(\tilde{\bx}_1,\by_1)\in \calT_{P_{\hat{X}_1\hat{Y}_1}}\big]\times\nonumber\\
&\quad\quad\quad\quad\quad
\mbox{Pr}
\big[N_{\by_1}(\hat{P}') \ge 1 , N_{\tilde{\bx}_1,\by_1}(\hat{P}) \ge 1 \big]
\label{eqn:case2bii}
\end{align}
To bound $\mbox{Pr}\big[N_{\by_1}(\hat{P}') \ge 1, N_{\tilde{\bx}_1,\by_1}(\hat{P}) 
\ge 1 \big]$, we consider two cases:\\

The first case is when $P_{\hat{X}_2\hat{Y}_1}=P_{\hat{X}_2'\hat{Y}_1'}$: 
in this case, $\big\{N_{\tilde{\bx}_1,\by_1}(\hat{P}) \ge 1\big\} 
\Rightarrow \big\{N_{\by_1}(\hat{P}') \ge 1 \big\}$. Therefore,
\begin{align}
\mbox{Pr}\big[N_{\by_1}(\hat{P}') \ge 1, 
N_{\tilde{\bx}_1,\by_1}(\hat{P}) \ge 1 \big] 
=& \mbox{Pr}\big[N_{\tilde{\bx}_1,\by_1}(\hat{P}) \ge 1 \big]\nonumber\\
\stackrel{.}{\le} & e^{n(R_2-\Iyxz)} \nonumber.
\end{align}
Replacing in (\ref{eqn:case2bii}), we get:
%%%%%%%%%%%%%%%%%%%%%%%%%%%%%
%[CAN STATE RANGES AND CONDITIONS]
%%%%%%%%%%%%%%%%%%%%%%%%%%%%%
\begin{align}
\bE_{\bX_1}\bE_{\calC_2}&
[N_{\bX_1,\by_1}^{\overline{\rho\lambda}}(\hat{P})
N_{\by_1}^{\rho}(\hat{P}')]\nonumber\\
& \stackrel{.}{\le}
\exp\big\{n \big[-\Ixz + R_2 - \Iyxz \big] \big\}. \label{eqn:Ecy2bba}
\end{align}

The other case is $P_{\hat{X}_2\hat{Y}_1}\ne P_{\hat{X}_2'\hat{Y}_1'}$: 
in this case, the same codeword $\bx_2$ 
cannot simultaneously satisfy $(\tilde{\bx}_1,\bx_2,\by_1)\in \calT_{\tp}$ 
and $(\bx_2,\by_1)\in \calT_{P_{\hat{X}_2'\hat{Y}_1'}}$. 
Therefore, we have that
\begin{align}
\mbox{Pr}\big[N_{\by_1}(\hat{P}') \ge 1,\quad &
N_{\tilde{\bx}_1,\by_1}(\hat{P}) \ge 1 \big]\nonumber\\
=& \mbox{Pr}\big[\exists \bx_2'\ne\bx_2: 
(\tilde{\bx}_1,\bx_2',\by_1)\in \calT_{\tp}, \nonumber\\
&\quad\quad\quad\quad\quad\quad\quad\quad
(\bx_2,\by_1)\in \calT_{P_{\hat{X}'_2,\hat{Y}_1'}} \big]\nonumber\\
\le&  \sum_{\bx_2\in\calC_2} 
\sum_{\bx_2'\ne\bx_2}\mbox{Pr}\big[(\tilde{\bx}_1,\bx_2',\by_1)\in\calT_{\tp},
\nonumber\\
& \quad\quad\quad\quad\quad\quad\quad\quad
(\bx_2,\by_1)\in \calT_{P_{\hat{X}_2',\hat{Y}_1'}} \big] \nonumber\\
\stackrel{.}{\le}& e^{n 2 R_2} e^{-n \Iyxz} e^{-n \Iyzp}. \nonumber
\end{align}
Replacing in (\ref{eqn:case2bii}), we get:
%%%%%%%%%%%%%%%%%%%%%%%%%%%%%
%[CAN STATE RANGES AND CONDITIONS]
%%%%%%%%%%%%%%%%%%%%%%%%%%%%%
\begin{align}
\bE_{\bX_1}\bE_{\calC_2}&
[N_{\bX_1,\by_1}^{\overline{\rho\lambda}}(\hat{P})
N_{\by_1}^{\rho}(\hat{P}')]\nonumber\\
\stackrel{.}{\le}& \exp\big\{n \big[-\Ixz+R_2-\Iyxz\nonumber\\
&\quad\quad\quad\quad\quad\quad\quad\quad+R_2-\Iyzp\big] \big\}.
\label{eqn:Ecy2bbb}
\end{align}
This completes the decomposition of $ \bE_{\calC_2} $ into the various
subcases.

\vspace{0.05in}

\noindent{\bf Consolidation}.
Next, we carry out a consolidation process that merges all of the above
subcases into a more compact expression, leading ultimately to the
expression in Theorem~\ref{thm:1}. Figure \ref{fig:consolidation} gives a schematic representation, in terms of a tree, of the various consolidation steps described below. The consolidation of (\ref{eqn:Ecy1a}) and (\ref{eqn:Ecy1b}) into (\ref{eqn:Ecy1}) was done before, but we include it in Fig. \ref{fig:consolidation} for completeness. Referring to Fig. \ref{fig:consolidation}, the consolidation starts at the deepest leaves of the tree and works its way up the nodes until it reaches the root.

We begin with the last set of subsubcases derived,
$ R_2 \geq \Iyxz $ and $ R_2 < \Iyxz
$ (expressions (\ref{eqn:Ecy2ba}), (\ref{eqn:Ecy2bba}), and
(\ref{eqn:Ecy2bbb}))
for the subcase $ R_2 < \Iyzp $, and consolidate them as follows:
\begin{align}
\bE_{\bX_1}\bE_{\calC_2} & \stackrel{.}{\le}
\exp\Bigg\{n\Big\{
1(R_2 \geq \Iyxz) \times \nonumber \\
& \big[\overline{\rho\lambda}(R_2-\Iyxz)-
\Ixz \nonumber\\
& \quad\quad\quad\quad\quad\quad \quad\quad\quad\quad\quad\quad +
R_2-\Iyzp \big] \nonumber \\
& + 1(R_2 < \Iyxz)1(P_{\hat{X}_2\hat{Y}_1}\ne
P_{\hat{X}_2'\hat{Y}_1'})\times \nonumber \\
& \quad \big[-\Ixz+R_2-\Iyxz\nonumber\\
&\quad\quad\quad\quad\quad\quad\quad\quad+R_2-\Iyzp\big] \nonumber \\
& + 1(R_2 < \Iyxz)1(P_{\hat{X}_2\hat{Y}_1}=
P_{\hat{X}_2'\hat{Y}_1'})\times \nonumber \\
& \quad \big[-\Ixz + R_2 - \Iyxz \big]
\Big\}\Bigg\}. \label{eqn:Ecy2bcomp} 
\end{align}
Next we would like to decompose the indicator 
$ 1(R_2 \geq \Iyxz) $ appearing in the initial part of this expression as 
\begin{align*}
1(R_2 & \geq \Iyxz) \\
 = & 1(R_2 \geq \Iyxz)
1(P_{\hat{X}_2\hat{Y}_1}=
P_{\hat{X}_2'\hat{Y}_1'})+ \\
& 
1(R_2 \geq \Iyxz)
1(P_{\hat{X}_2\hat{Y}_1}\ne
P_{\hat{X}_2'\hat{Y}_1'}) \\
 = & 1(R_2 \geq \Iyxz)
1(P_{\hat{X}_2\hat{Y}_1}\ne
P_{\hat{X}_2'\hat{Y}_1'}), \\
\end{align*}
where we are taking into account in the last step that for the present
subcase $ R_2 < \Iyzp $, 
$ 1(R_2 \geq \Iyxz)
1(P_{\hat{X}_2\hat{Y}_1}=
P_{\hat{X}_2'\hat{Y}_1'}) = 0 $ since for
$ P_{\hat{X}_2\hat{Y}_1}= P_{\hat{X}_2'\hat{Y}_1'} $ we have
$ R_2 < \Iyzp = \Iyz \leq \Iyxz $.

Applying this decomposition to (\ref{eqn:Ecy2bcomp}), then
combining terms having the same
indicators $ 1(P_{\hat{X}_2\hat{Y}_1}\ne
P_{\hat{X}_2'\hat{Y}_1'}) $, and $  1(P_{\hat{X}_2\hat{Y}_1}=
P_{\hat{X}_2'\hat{Y}_1'}) $, and replacing indicators by $
\min\{\cdots\} $ as appropriate (similar to (\ref{eqn:Ecy1})),
we simplify~(\ref{eqn:Ecy2bcomp}) to
\begin{align}
\bE_{\bX_1}&\bE_{\calC_2} \nonumber \\
\stackrel{.}{\le} &
\exp\Bigg\{n\Big\{
1(P_{\hat{X}_2\hat{Y}_1}\ne
P_{\hat{X}_2'\hat{Y}_1'})\big[-\Ixz+ \nonumber \\
& \quad \min\{
\overline{\rho\lambda}(R_2{-}\Iyxz),R_2{-}\Iyxz\} \nonumber\\
&\quad\quad+R_2-\Iyzp\big] \nonumber \\
& + 1(P_{\hat{X}_2\hat{Y}_1}=
P_{\hat{X}_2'\hat{Y}_1'})1(R_2 < \Iyxz)\times \nonumber \\
& \quad \big[-\Ixz + R_2 - \Iyxz \big]\Big] 
\Big\}\Bigg\} .\label{eqn:Ecy2b}
\end{align}
This is valid for the subcase
$ R_2 < \Iyzp $.

Next, we consolidate~(\ref{eqn:Ecy2a}) from the subcase $ R_2 \geq
\Iyzp $ with~(\ref{eqn:Ecy2b}) and insert the result
into~(\ref{eqn:Ecy2outer}) to get
\begin{align}
\bE_{\calC_2} & \stackrel{.}{\le}
\exp\Bigg\{n\Big\{ -\rho\Ixyzp \nonumber\\
&+1(R_2 {\geq} \Iyzp)\Big[{-}\Ixz{+}\rho (R_2{-}\Iyzp) \nonumber \\
& {+}\min\{\overline{\rho\lambda}(R_2{-}\Iyxz),(R_2{-}\Iyxz)\} \Big]
\nonumber  \\ 
& {+} 1(R_2 {<} \Iyzp) \Big[1(P_{\hat{X}_2\hat{Y}_1}{\ne}
P_{\hat{X}_2'\hat{Y}_1'})\big[-\Ixz \nonumber \\
& {+}\min\{
\overline{\rho\lambda}(R_2{-}\Iyxz),R_2{-}\Iyxz\} \nonumber\\
&\quad\quad\quad+R_2-\Iyzp\big] \nonumber \\
& \quad + 1(P_{\hat{X}_2\hat{Y}_1}=
P_{\hat{X}_2'\hat{Y}_1'})1(R_2 < \Iyxz)\times \nonumber \\
& \quad\quad \big[-\Ixz + R_2 - \Iyxz \big]\Big] \Big\}\Bigg\},
  \label{eqn:Ecy2comp1}  
\end{align}
which applies to the range $ R_2 < \Iyxzp $.
Again, expanding all terms against the 
indicators $ 1(P_{\hat{X}_2\hat{Y}_1}\ne
P_{\hat{X}_2'\hat{Y}_1'}) $, and $  1(P_{\hat{X}_2\hat{Y}_1}=
P_{\hat{X}_2'\hat{Y}_1'}) $, and, as above, replacing indicators by $
\min\{\cdots\} $ as appropriate, we obtain
\begin{align}
\bE_{\calC_2} & \stackrel{.}{\le}
\exp\Bigg\{n\Big\{1(P_{\hat{X}_2\hat{Y}_1}\ne
P_{\hat{X}_2'\hat{Y}_1'})\Big[ -\rho\Ixyzp \nonumber \\
& \quad -\Ixz+\min\{
\overline{\rho\lambda}(R_2-\Iyxz), \nonumber \\
& \quad\quad\quad\quad\quad R_2-\Iyxz\} \nonumber\\
& \quad +\min\{\rho(R_2-\Iyzp),R_2-\Iyzp\}\Big] \nonumber \\
& 1(P_{\hat{X}_2\hat{Y}_1}=
P_{\hat{X}_2'\hat{Y}_1'}) \times \nonumber \\
&\quad\Big[ -\rho\Ixyzp +1(R_2 \geq \Iyz)\times \nonumber \\
&\quad\quad \big[-\Ixz + \rho (R_2-\Iyzp) \nonumber \\
& \quad\quad +\min\{\overline{\rho\lambda}(R_2-\Iyxz),\nonumber\\
& \quad\quad\quad\quad R_2-\Iyxz\} \big] + 1(R_2 < \Iyz)\times \nonumber \\
&\quad\quad \big[-\Ixz + R_2 - \Iyxz \big]\Big] \Big\}\Bigg\}.
\label{eqn:Ecy2comp2} 
\end{align}
Using the identity (proved via the chain rule)
\begin{multline*}
\Ixyzp + \Iyzp = \Iyxzp + \Ixzp 
\end{multline*}
twice, we can rewrite the term
\begin{multline*}
-\rho\Ixyzp +  \min\{\rho(R_2-\Iyzp),\\R_2-\Iyzp\}
\end{multline*}
appearing after the indicator
$ 1(P_{\hat{X}_2\hat{Y}_1}\ne P_{\hat{X}_2'\hat{Y}_1'}) $
in~(\ref{eqn:Ecy2comp2}) as
\begin{multline*}
-\rho\Ixzp + \min\{\rho(R_2-\Iyxzp),\\R_2-\overline{\rho}\Iyzp-\rho\Iyxzp\}.
\end{multline*}
Similarly, we can decompose the term $ -\rho \Ixyzp $ appearing after
the indicator $ 1(P_{\hat{X}_2\hat{Y}_1}= P_{\hat{X}_2'\hat{Y}_1'}) $ 
against the indicators $ 1(R_2 \geq \Iyz $ and $ 1(R_2 < \Iyz) $, and
use the above identity to combine it with $ \rho(R_2 - \Iyzp) $
appearing after the indicator $ 1(R_2 \geq \Iyz) $.
Incorporating these steps, we can rewrite~(\ref{eqn:Ecy2comp2}) as
\begin{align}
\bE_{\calC_2} & \stackrel{.}{\le}
\exp\Bigg\{n\Big\{ 
1(P_{\hat{X}_2\hat{Y}_1}{\ne}
P_{\hat{X}_2'\hat{Y}_1'}) \Big[ {-}\Ixz {-} \rho\Ixzp \nonumber \\
& \quad +\min\{
\overline{\rho\lambda}(R_2-\Iyxz), R_2-\Iyxz\}\nonumber\\
& \quad+\min\{
R_2-\overline{\rho}\Iyzp-\rho\Iyxzp, \nonumber \\
& \quad\quad\quad\quad\quad\quad \rho(R_2-\Iyxzp)
\}\Big] \nonumber \\
& +1(P_{\hat{X}_2\hat{Y}_1}=
P_{\hat{X}_2'\hat{Y}_1'}) \times \nonumber \\
&\quad\Big[1(R_2 \geq \Iyz)\big[-\Ixz-\rho\Ixzp\nonumber \\
& \quad\quad {+}\min\{\overline{\rho\lambda}(R_2{-}\Iyxz),R_2{-}\Iyxz\}
    \nonumber \\
&\quad\quad {+} \rho (R_2-\Iyxzp)\big] \nonumber \\
&\quad+ 1(R_2 < \Iyz)\big[-\Ixz + R_2 \nonumber \\
&\quad\quad - \Iyxz -\rho\Ixyzp\big]\Big]\Big\}\Bigg\}.
\label{eqn:Ecy2} 
\end{align}

Finally, we consolidate~(\ref{eqn:Ecy1}) from the range
$ R_2 \geq \Iyxzp $ with the just obtained (\ref{eqn:Ecy2}) 
(for the range $ R_2 < \Iyxzp $) to get
\begin{align}
\bE_{\calC_2} & \stackrel{.}{\le}
\exp\Bigg\{n\Big\{1(R_2 \geq \Iyxzp)\times \nonumber \\
& \Big[-I(\hat{X}_1;\hat{Y}_1) - \rho \Ixzp \nonumber\\
& {+}\min\{\overline{\rho\lambda}(R_2 {-} \Iyxz), (R_2{-}\Iyxz)\} \nonumber \\
&{+} \rho\lambda(R_2-\Iyxzp)\Big]
\nonumber \\
& + 1(R_2 < \Iyxzp)\Big[
1(P_{\hat{X}_2\hat{Y}_1}{\ne}
P_{\hat{X}_2'\hat{Y}_1'})\times \nonumber \\
&\quad \Big[ {-}\Ixz {-} \rho\Ixzp \nonumber \\
& \quad +\min\{
\overline{\rho\lambda}(R_2{-}\Iyxz), R_2{-}\Iyxz\}\nonumber\\
& \quad+\min\{
R_2-\overline{\rho}\Iyzp-\rho\Iyxzp, \nonumber \\
& \quad\quad\quad\quad\quad\quad\rho(R_2-\Iyxzp)\}
\Big] \nonumber \\
& +1(P_{\hat{X}_2\hat{Y}_1}=
P_{\hat{X}_2'\hat{Y}_1'}) \times \nonumber \\
&\quad\Big[1(R_2 \geq \Iyz)\big[-\Ixz-\rho\Ixzp\nonumber \\
& \quad {+}\min\{\overline{\rho\lambda}(R_2{-}\Iyxz),R_2{-}\Iyxz\}
    \nonumber \\
&\quad {+} \rho (R_2-\Iyxzp)\big] \nonumber \\
&\quad+ 1(R_2 < \Iyz)\big[-\Ixz + R_2 \nonumber \\
&\quad\quad - \Iyxz -\rho\Ixyzp\big]\Big]
\Big]\Big\}\Bigg\}.
\label{eqn:Ecycomp} 
\end{align}

As before, after expanding the first indicator $ 1(R_2 \geq \Iyxzp) $
against $ 1(P_{\hat{X}_2\hat{Y}_1}\ne
P_{\hat{X}_2'\hat{Y}_1'}) $, and $  1(P_{\hat{X}_2\hat{Y}_1}=
P_{\hat{X}_2'\hat{Y}_1'}) $, and combining terms, we obtain
\begin{align}
\bE_{\calC_2} & \stackrel{.}{\le}
\exp\Bigg\{n\Big\{
1(P_{\hat{X}_2\hat{Y}_1}{\ne}
P_{\hat{X}_2'\hat{Y}_1'})\Big[ {-}\Ixz {-} \rho\Ixzp \nonumber \\
& \quad +\min\{
\overline{\rho\lambda}(R_2{-}\Iyxz), R_2{-}\Iyxz\}\nonumber\\
& \quad +\min\{
R_2-\overline{\rho}\Iyzp -\rho\Iyxzp, \nonumber \\
& \quad\quad
\rho(R_2-\Iyxzp),\rho\lambda(R_2-\Iyxzp)\}\Big] \nonumber \\
& +1(P_{\hat{X}_2\hat{Y}_1}=
P_{\hat{X}_2'\hat{Y}_1'}) \times \nonumber \\
&\quad\Big[1(R_2 \geq \Iyz)\big[-\Ixz-\rho\Ixzp\nonumber \\
& \quad {+}\min\{\overline{\rho\lambda}(R_2{-}\Iyxz),R_2{-}\Iyxz\}
  \nonumber \\
&\quad {+} \min\{\rho (R_2{-}\Iyxzp),\rho\lambda(R_2{-}\Iyxzp)\}
\big] \nonumber \\
&\quad+ 1(R_2 < \Iyz)\big[-\Ixz + R_2 \nonumber \\
&\quad\quad - \Iyxz -\rho\Ixyzp\big]
\Big]\Big\}\Bigg\},
\label{eqn:Ecy} 
\end{align}
where, in simplifying, we have made use of the identity
\begin{align*}
1&(R_2 \geq \Iyxzp)\rho\lambda(R_2-\Iyxzp)+ \\
& \quad 1(R_2 < \Iyxzp)\min\{ R_2-\overline{\rho}\Iyzp \\
& -\rho\Iyxzp, \rho(R_2-\Iyxzp)\} \\
& = \min\{
R_2-\overline{\rho}\Iyzp-\rho\Iyxzp,\\
& \quad \rho(R_2-\Iyxzp), \rho\lambda(R_2-\Iyxzp)\},
\end{align*}
along with
\begin{align*}
&1(P_{\hat{X}_2\hat{Y}_1}=
P_{\hat{X}_2'\hat{Y}_1'}) 1(R_2 \geq \Iyxzp) \\
& = 1(P_{\hat{X}_2\hat{Y}_1}{=}
P_{\hat{X}_2'\hat{Y}_1'}) 1(R_2 {\geq} \Iyz)1(R_2 {\geq} \Iyxzp),
\end{align*}
and finally
\begin{align*}
1&(R_2 \geq \Iyxzp)\rho\lambda(R_2-\Iyxzp)+ \\
& 1(R_2 < \Iyxzp)\rho(R_2-\Iyxzp) \\
& = \min\{\rho(R_2{-}\Iyxzp),\rho\lambda(R_2{-}\Iyxzp)\}.
\end{align*}

We use (\ref{eqn:Ecy}) in (\ref{eqn:EcyAB}), 
add over all vectors $\by_1$, decompose all joint-type-dependent terms
appearing in (\ref{eqn:EcyAB}), as well as the term $ nH(\hat{Y}_1) $
arising from the summation over $ \by_1$ per type, against the indicators
$ 1(P_{\hat{X}_2\hat{Y}_1}\ne P_{\hat{X}_2'\hat{Y}_1'}) $ and $
1(P_{\hat{X}_2\hat{Y}_1}=P_{\hat{X}_2'\hat{Y}_1'}) $, 
and finally optimize over the types $\tp$, $\tpp$
to obtain:
\begin{align}
\bE_{\calC_1,\calC_2}&(P_{E_1}) \stackrel{.}{\le} 
\exp\Bigg\{n\Bigg\{
-R_2+\rho R_1 + \max\Bigg\{ \nonumber\\
&  \max_{\substack{\tp,\tpp\\
P_{\hat{X}_1}=P_{\hat{X}_1'}=Q_1, \\P_{\hat{X}_2}=P_{\hat{X}_2'}=Q_2, 
\\P_{\hat{Y}}=P_{\hat{Y}_1'}
\\P_{\hat{X}_2\hat{Y}_1}\ne P_{\hat{X}_2'\hat{Y}_1'}, }}
\Bigg[ \overline{\rho\lambda}\bE_{\hat{X}_1\hat{X}_2\hat{Y}_1}\log 
q_1(\hat{Y}_1|\hat{X}_1,\hat{X}_2) \nonumber \\
& +\rho\lambda\bE_{\hat{X}_1'\hat{X}_2'\hat{Y}_1'}
\log q_1(\hat{Y}_1'|\hat{X}_1',\hat{X}_2')\nonumber\\
&+H(\hat{Y}_1|\hat{X}_1){-} \rho\Ixzp \nonumber \\
& \quad +\min\{
\overline{\rho\lambda}(R_2{-}\Iyxz), R_2{-}\Iyxz\}\nonumber\\
& \quad+\min\{
R_2-\overline{\rho}\Iyzp -\rho\Iyxzp, \nonumber \\
& \quad\quad\rho(R_2{-}\Iyxzp),\rho\lambda(R_2{-}\Iyxzp)\}
\Bigg];\nonumber\\
&\max_{\substack{\tp,\tpp\\
P_{\hat{X}_1}=P_{\hat{X}_1'}=Q_1, 
\\P_{\hat{X}_2}=P_{\hat{X}_2'}=Q_2, 
\\P_{\hat{X}_2\hat{Y}_1}=P_{\hat{X}_2'\hat{Y}_1'}}} 
\Bigg[\overline{\rho\lambda}\bE_{\hat{X}_1\hat{X}_2\hat{Y}_1}
\log q_1(\hat{Y}_1|\hat{X}_1,\hat{X}_2) \nonumber \\
&+\rho\lambda\bE_{\hat{X}_1'\hat{X}_2'
\hat{Y}_1'}\log q_1(\hat{Y}_1'|\hat{X}_1',\hat{X}_2')\nonumber\\
&+1(R_2 \geq \Iyz)\big[H(\hat{Y}_1|\hat{X}_1)-\rho\Ixzp\nonumber \\
& \quad {+}\min\{\overline{\rho\lambda}(R_2{-}\Iyxz),R_2{-}\Iyxz\}
  \nonumber \\
&\quad {+} \min\{\rho (R_2-\Iyxzp),\nonumber \\ 
& \quad\quad \rho\lambda(R_2-\Iyxzp)\}
\big] \nonumber \\
&\quad+ 1(R_2 < \Iyz)\big[H(\hat{Y}_1|\hat{X}_1) + R_2 \nonumber \\
&\quad\quad - \Iyxz -\rho\Ixyzp\big]\Bigg]\Bigg\}\Bigg\}\Bigg\}
\label{eqn:e9}
\end{align}
Note that the term $ H(\hat{Y}_1) $ mentioned above has been combined
with the term $ -\Ixy $ appearing in all subcases of~(\ref{eqn:Ecy}) to 
yield the $ H(\hat{Y}_1|\hat{X}_1) $ appearing throughout~(\ref{eqn:e9}).

The expression in Theorem~\ref{thm:1} is obtained from (\ref{eqn:e9})
by dropping the constraint $  P_{\hat{X}_2\hat{Y}_1}\ne
P_{\hat{X}_2'\hat{Y}_1'}, $ from the first maximization (which, given
the continuity of the underlying terms, is not really a constraint
anyway), by noting that if, in the resulting expression, the second
maximization is attained 
when $ R_2 \geq \Iyz $, it will be dominated by the first maximization
so that the second maximization can be restricted to the case $ R_2 < \Iyz $,
and finally by negating the resulting exponent (and propagating the
negation as $ -\max\{\cdots\} = \min\{-\cdots\} $ throughout).

\section{A Lower Bound to $ E_{R,1}$}
\label{sec:altexpr}
\renewcommand{\theequation}{\thesection.\arabic{equation}}
\setcounter{equation}{0}

We can lower bound the maximization of (\ref{eqn:erx}) over $\rho$ and
$\lambda$ by applying the min-max
theorem twice, as follows.  

First we introduce a new parameter
$\theta$ and bound (\ref{eqn:erx}) as
\begin{align}
E&_{R,1} \geq \min_{\theta \in [0,1]}
\Bigg\{
R_2-\rho R_1 + \theta \times \\
&\min_{\substack{(\tpa{1},\\\tppa{1})\\
\in \calS_1(Q_1,Q_2)}}
f_1\left(\rho,\lambda,\tpa{1},\tppa{1}\right) + \nonumber\\
& \overline{\theta}\min_{\substack{(\tpa{2},\\\tppa{2})\\
\in \calS_2(Q_1,Q_2)}} 
 f_2\left(\rho,\lambda,\tpa{2},\tppa{2}\right)\Bigg\}
\label{eqn:erx2}
\end{align}
where $ \overline{\theta}=1-\theta$ and we have dropped the constraint involving $R_2$ from $\calS_2$, resulting in a lower bound, and making $\calS_2$ convex.

Letting $\gamma = \rho \lambda $, we claim that for fixed $\theta$,
the expression in (\ref{eqn:erx2}) being minimized over $\theta$ above is convex in $(\rho,\gamma)$.  This follows from the fact that for fixed $
\tpa{1}, $ $ \tppa{1}, $ $ \tpa{2}, $ $ \tppa{2})$, both $f_1$ and $f_2$ are affine
in $(\rho,\gamma)$.  The only problem would come from the $\max$'s
appearing in these expressions, but it can be checked that these
maximizations are independent of $(\rho,\gamma)$ for fixed $
(\tpa{1},$ $ \tppa{1}, $ $ \tpa{2}, $ $ \tppa{2})$.  Letting $\Sigma = \{(x,y):x\in
[0,1],y\in[0,x]\}$, we can thus apply the min-max
theorem of convex analysis (twice) as follows
\begin{align}
&E^*_{R,1} \nonumber \\
&\geq \max_{(\rho,\gamma)\in\Sigma} \min_{\theta \in [0,1]}
\Bigg\{
R_2-\rho R_1 + \theta \times \nonumber \\
&\min_{\substack{(\tpa{1},\\\tppa{1})\\
\in \calS_1(Q_1,Q_2)}}
f_1\left(\rho,\gamma,\tpa{1},\tppa{1}\right) + \nonumber \\
&\overline{\theta}\min_{\substack{(\tpa{2},\\\tppa{2})\\
\in \calS_2(Q_1,Q_2)}} 
 f_2\left(\rho,\gamma,\tpa{2},\tppa{2}\right)\Bigg\} \nonumber \\
&=   \min_{\theta \in [0,1]} \max_{(\rho,\gamma)\in\Sigma}
\Bigg\{
R_2-\rho R_1 + \theta \times \nonumber \\
&\min_{\substack{(\tpa{1},\\\tppa{1})\\
\in \calS_1(Q_1,Q_2)}}
f_1\left(\rho,\gamma,\tpa{1},\tppa{1}\right) + \nonumber \\
& \overline{\theta}\min_{\substack{(\tpa{2},\\\tppa{2})\\
\in \calS_2(Q_1,Q_2)}} 
 f_2\left(\rho,\gamma,\tpa{2},\tppa{2}\right)\Bigg\} \nonumber \\
&=   \min_{\theta \in [0,1]} \max_{(\rho,\gamma)\in\Sigma}
\min_{\substack{(\tpa{1},\tppa{1},\\\tpa{2},\tppa{2})\\\in
    \calS_1(Q_1,Q_2)\times \calS_2(Q_1,Q_2)}} \Bigg\{ 
R_2-\rho R_1 + \nonumber \\
& \quad\quad\quad\theta f_1\left(\rho,\gamma,\tpa{1},\tppa{1}\right) + \nonumber \\
& \quad\quad\quad\overline{\theta} f_2\left(\rho,\gamma,\tpa{2},\tppa{2}\right)\Bigg\}\nonumber \\
&= \min_{\theta \in [0,1]}
\min_{\substack{(\tpa{1},\tppa{1},\\\tpa{2},\tppa{2})\\\in
    \calS_1(Q_1,Q_2)\times \calS_2(Q_1,Q_2)}} \max_{(\rho,\gamma)\in\Sigma}
\Bigg\{ 
R_2-\rho R_1 + \nonumber \\
&\quad\quad\quad \theta f_1\left(\rho,\gamma,\tpa{1},\tppa{1}\right) + \nonumber \\
&\quad\quad\quad \overline{\theta} f_2\left(\rho,\gamma,\tpa{2},\tppa{2}\right)\Bigg\}\label{eq:innermax}
\end{align}
Since, as noted above, for fixed ($
\theta,$ $\tpa{1},$ $\tppa{1},$ $\tpa{2},$ $\tppa{2}$), both $f_1$ and $f_2$ are  
affine in $(\rho,\gamma)$, the inner maximization
in~(\ref{eq:innermax}) is attained at one of the points $
(\rho,\gamma) = \{(0,0),(1,0),(1,1)\}$.  After simplification, we
obtain
\begin{align*}
&E_{R,1}^* \geq \min_{\theta \in [0,1]} 
\min_{\substack{(\tpa{1},\tppa{1},\\\tpa{2},\tppa{2})\\\in 
    \calS_1(Q_1,Q_2)\times \calS_2(Q_1,Q_2)}} \max\Bigg\{ \\
& \theta \Big[-\elqa{1}-\Hzxa{1}+\\
& \quad\quad\Iyxza{1}+ |\Iyzpa{1}-R_2|^+\Big] + \\
& \overline{\theta}\Big[-\elqa{2}-\Hzxa{2}+\\
& \quad\quad\Iyxza{2}\Big];\\ & \\
& -R_1+ \theta \Big[-\elqa{1}-\\
& \quad\quad\Hzxa{1}+\Ixzpa{1}+ \\
& \quad\quad\Iyxza{1}+|\Iyxzpa{1}-R_2|^+\Big] + \\
& \overline{\theta}\Big[-\elqa{2}-\Hzxa{2}+\\
& \quad\quad\Ixyzpa{2}+\Iyxza{2}\Big]; \\ & \\
& -R_1+ \theta \Big[-\elqpa{1}-\\
& \quad\quad\Hzxa{1}+\Ixzpa{1}+ \\
& \quad\quad\Iyxzpa{1}+|\Iyxza{1}-R_2|^+\Big] + \\
& \overline{\theta}\Big[-\elqpa{2}-\Hzxa{2}+\\
& \quad\quad\Ixyzpa{2}+\Iyxza{2}\Big]\Bigg\}
\end{align*}

Next, we note the identities 
\begin{align*}
&\Iyxz = \Ixy + \Hzx - \Hzxy \\
&\Ixyz = \Ixy + \Hzy - \Hzxy \\
&\Dpzxyq = -\Hzxy-\\
&\quad\quad\quad\elq 
\end{align*}
and use them, with the shorthand
$ D^{(m)} = \Da{m} $ and $ D^{'(m)} = \Dpa{m} $, for $ m \in \{ 1,2 \}$,
 to rewrite the bound as
\begin{align}
&E_{R,1}^* \geq \min_{\theta \in [0,1]}
\min_{\substack{(\tpa{1},\tppa{1},\\\tpa{2},\tppa{2})\\\in 
    \calS_1(Q_1,Q_2)\times \calS_2(Q_1,Q_2)}} \max\Bigg\{ \nonumber \\
& \theta \Big[D^{(1)}+\Ixya{1}+|\Iyzpa{1}-R_2|^+\Big] + \nonumber \\
& \overline{\theta}\Big[D^{(2)}+\Ixya{2}\Big];\nonumber\\\nonumber \\
& -R_1+ \theta \Big[D^{(1)}+\Ixya{1}+\Ixzpa{1}\nonumber\\
& \quad\quad\quad +|\Iyxzpa{1}-R_2|^+\Big] + \nonumber\\
& \overline{\theta}\Big[D^{(2)}+\Ixya{2}+\Ixyzpa{2}\Big]; \nonumber\\\nonumber\\
& -R_1+ \theta \Big[D^{'(1)}+\Ixypa{1}+\Ixza{1}+\nonumber\\
& \quad\quad\quad|\Iyxza{1}-R_2|^+\Big] + \nonumber\\
& \overline{\theta}\Big[D^{'(2)}+\Ixypa{2}+\Ixyza{2}\Big]\Bigg\}
\label{eq:altexprfinal}
\end{align}
where in simplifying the third expression in the maximization we have
also exploited the constraints $ H(\hat{Y}_1^{(1)}) = H(\hat{Y}_1^{'(1)}) $
and  $ H(\hat{Y}_1^{(2)}|\hat{X}_2^{(2)}) =
H(\hat{Y}_1^{'(2)}|\hat{X}_2^{'(2)}) $.

For $ R_1 = 0 $ we can further simplify this expression.  In
particular, for $ R_1 = 0 $, the first term in the inner maximization
is readily seen to be always smaller than the second term.
Additionally, the second and third terms are symmetric in the primed
and non-primed joint distributions, which, together with the readily
established joint convexity
of the maximum of these two terms on the constraint set, imply that the inner
minimization over the joint types is achieved when the primed and
non-primed joint distributions are equal, in which case the two terms are equal. 
Therefore, at $ R_1 = 0 $
we have
\begin{align}
E_{R,1}^* & \geq \min_{\theta \in [0,1]}
\min_{\substack{(\tpa{1},\tpa{2}): \\
P_{\hat{X}^{(1)}_1}=P_{\hat{X}^{(2)}_1}=
Q_1, P_{\hat{X}^{(1)}_2}=P_{\hat{X}^{(2)}_2}=
Q_2}} \nonumber \\
& \theta \Big[D^{(1)}+\Ixya{1}+ \nonumber \\
& \quad\quad\Ixza{1}+|\Iyxza{1}-R_2|^+\Big] + \nonumber\\
& \overline{\theta}\Big[D^{(2)}+\Ixya{2}+\Ixyza{2}\Big]
\end{align}
or
\begin{align}
E_{R,1}^* &\geq \min\Bigg\{ \min_{\substack{\tp:\\P_{\hat{X}_1}=Q_1, P_{\hat{X}_2}=Q_2}}
 \Big[D+\Ixy+\Ixz\nonumber\\
& \quad\quad\quad +|\Iyxz-R_2|^+\Big];\nonumber\\
& \nonumber\\
& \min_{\substack{\tp:\\P_{\hat{X}_1}=Q_1, P_{\hat{X}_2}=Q_2}} \Big[D+\Ixy+\Ixyz\Big]\Bigg\}
\end{align}
where $ D = \Dpzxyq $.

Simplifying $ E_{B,1} $ at $ R_1 = 0 $ gives
\begin{align}
E_{B,1} &= \max\Bigg\{\min_{\substack{\tp:\\P_{\hat{X}_1}=Q_1, P_{\hat{X}_2}=Q_2}}
 \Big[D+\Ixy + \Ixz\Big];\nonumber \\
&\quad\min \Bigg\{\min_{\substack{\tp:\\P_{\hat{X}_1}=Q_1, P_{\hat{X}_2}=Q_2}}
 \Big[D+\Ixy+ \nonumber \\
& \quad\quad\quad +|\Ixz+\Iyxz-R_2|^+\Big];\nonumber\\ 
& \nonumber\\
& \quad\min_{\substack{\tp:\\P_{\hat{X}_1}=Q_1, P_{\hat{X}_2}=Q_2}}
 \Big[D+\Ixy+\Ixyz\Big] \Bigg\} \Bigg\}
\end{align}
which is seen to be no bigger than the above lower bound on $ E^*_{R,1} $, since
$ |\Iyxz-R_2|^+ \geq 0 $, $\Ixyz \geq \Ixz $, and $ \Ixz+|\Iyxz-R_2|^+ \geq
|\Ixz+\Iyxz-R_2|^+ $.

Another application of the lower bound (\ref{eq:altexprfinal}) is in
determining the set of rate pairs $ R_1,R_2 $ for which $ E_{R,1}^* >
0 $. 
Let $ (\hat{X}_1,\hat{X}_2) $ be independent with marginal distributions
$ Q_1 $ and $ Q_2 $ and $ \hat{Y}_1 $ be the result of 
$ (\hat{X}_1,\hat{X}_2) $ passing through the channel $ q_1 $.  We
shall argue that if $ R_1 < \Ixz+|\Iyxz-R_2|^+ $ $ = \Ixz+|
I(\hat{X}_2;\hat{Y}_1|\hat{X}_1)-R_2|^+ $.
 and $ R_1 < \Ixyz = I(\hat{X}_1;\hat{Y}_1|\hat{X}_2) $
then the expression (\ref{eq:altexprfinal}) must be greater than 0.
Indeed, for the expression (\ref{eq:altexprfinal}) to equal 0, we see from
the first term in the inner maximum that the 
minimizing $ \theta $ and joint distributions must satisfy one of the
following: 
case 1: $ \theta = 1 $, $ D^{(1)} = 0$, and $ \Ixya{1} = 0 $; case 2:
$ \theta = 0 $, $ D^{(2)} = 0$, and $ \Ixya{2} = 0 $; or case 3:
$ 0< \theta < 1 $, $ D^{(1)} = D^{(2)} = 0$, and $  \Ixya{1} =
\Ixya{2} = 0 $.  If case 1 holds then $
(\hat{X}_1^{(1)},\hat{X}_2^{(1)},\hat{Y}_1^{(1)}) $ 
necessarily have the same joint distribution as $
(\hat{X}_1,\hat{X}_2,\hat{Y}_1) $, in which case, we see  from the
third term in the maximum in~(\ref{eq:altexprfinal}) that $ 
R_1 \geq \Ixz+|\Iyxz-R_2|^+ $.  Similarly, if case 2 holds then it
follows that $ (\hat{X}_1^{(2)},\hat{X}_2^{(2)},\hat{Y}_1^{(2)}) $ 
have the same joint distribution as $
(\hat{X}_1,\hat{X}_2,\hat{Y}_1) $, in which case, it follows again
from the third term in the maximum that $ R_1 \geq \Ixyz $.  Finally,
if case 3 holds then both $
(\hat{X}_1^{(1)},\hat{X}_2^{(1)},\hat{Y}_1^{(1)}) $ and 
$ (\hat{X}_1^{(2)},\hat{X}_2^{(2)},\hat{Y}_1^{(2)}) $ have the same
distribution as $
(\hat{X}_1,\hat{X}_2,\hat{Y}_1) $, in which case, after writing $ R_1 =
\theta R_1 + \overline{\theta}R_1 $, we see again that either $ R_1 \geq
\Ixz+|\Iyxz-R_2|^+ $ or $ R_1 \geq \Ixyz $ must hold.  Thus, the three cases
together establish the above claim that if 
$ R_1 < \Ixz+|
I(\hat{X}_2;\hat{Y}_1|\hat{X}_1)-R_2|^+ $ and $ R_1 < 
I(\hat{X}_1;\hat{Y}_1|\hat{X}_2) $ 
then the expression (\ref{eq:altexprfinal}), and hence $ E^*_{R,1} $,
must be greater than 0.  It can be checked that this region is
equivalent to 
\begin{multline*}
\{ R_1 < \Ixz \} \cup \Big\{\{R_1 + R_2 <
I(\hat{Y}_1;\hat{X}_1,\hat{X}_2) 
\} \\ \cap \{R_1 < I(\hat{X}_1;\hat{Y}_1|\hat{X}_2)\Big\} 
\end{multline*}
which is represented in Fig. \ref{fig:rateregion} in Section \ref{sec:main}.
It is shown in~\cite{Cha08} that for 
the ensemble of constant composition codes comprised of i.i.d.\ codewords
uniformly distributed over the types $ Q_1 $ and $ Q_2 $,
the exponential decay rate of the average probability of error for user 1
must necessarily be zero for rate pairs outside of this region, even
for optimum, maximum likelihood decoding.

\end{document}